\documentclass[manuscript]{emulateapj} 
\shorttitle{Reionization powered by GPUs} 
\shortauthors{Aubert \& Teyssier} 
\usepackage{graphicx} 
\begin{document} 

\title{Reionization Simulations powered by GPUs I : On the structure of the Ultra-Violet radiation field}

\author{Dominique Aubert} \affil{Observatoire Astronomique, Universite de
  Strasbourg, UMR 7550, 11 rue de l'Universite, F-67000 Strasbourg, France} \and 
\author{Romain Teyssier\altaffilmark{1}} \affil{IRFU, CEA Saclay, Batiment 709, F-91191 Gif--sur--Yvette Cedex, France}

\altaffiltext{1}{Instit\"ut fur Theoretische Physik, Universit\"at Z\"urich, Winterthurerstrasse 190, 8057, Z\"urich, Switzerland}

\begin{abstract}
	We present a set of cosmological simulations with radiative transfer in order to model the reionization history of the universe from $z = 18$ down to $z=6$. Galaxy formation and the associated star formation are followed self-consistently with gas and dark matter dynamics using the {\tt RAMSES} code, while radiative transfer is performed as a post-processing step using a moment-based method with M1 closure relation in the {\tt ATON} code. 	
	
	 The latter has been ported to a multiple Graphical Processing Units (GPU) architecture using the CUDA language together with the MPI library, resulting in an overall acceleration that allows us to tackle radiative transfer problems at a significantly higher resolution than previously reported: $1024^3$ + 2 levels of refinement for the hydrodynamics adaptive grid and $1024^3$ for the radiative transfer Cartesian grid. We reach typical acceleration factor close to $100\times$ when compared to the CPU version, allowing us to perform 1/4 million time steps in less than 3000 GPU hours.
	
	We observe good convergence properties between our different resolution runs for various volume- and mass-averaged quantities such as neutral fraction, UV background and Thomson optical depth, as long as the effects of finite resolution on the star formation history are properly taken into account. We also show that the neutral fraction depends on the total mass density, in a way close to the predictions of photoionization equilibrium, as long as the effect of self-shielding are included in the background radiation model. Although our simulation suite has reached unprecedented mass and spatial resolution, we still fail at reproducing the $z\sim 6$ constraints on the neutral fraction of hydrogen and the intensity of the UV background.
	
	In order to account for unresolved density fluctuations, we have modified our chemistry solver with a simple clumping factor model. Using our most spatially resolved simulation (12.5 Mpc/h with 1024$^3$ particles) to calibrate our subgrid model, we have resimulated our largest box (100 Mpc/h with 1024$^3$  particles) with the modified chemistry, successfully reproducing the observed level of neutral Hydrogen in the spectra of high redshift quasars. We however didn't reproduce (by a factor of 2) the average photoionization rate inferred from the same observations.  We argue that this discrepancy could be partly explained by the fact that the average radiation intensity and the average neutral fraction depends on different regions of the gas density distribution, so that one quantity cannot be simply deduced from the other. 

\end{abstract}


\section{Introduction} 

After self-gravity, hydrodynamics and radiative cooling ( see e.g. \citet{1985ApJS...57..241E, 1991ApJS...75..231H, 1992ApJS...78..341C,1996ApJS..105...19K, 1998ARA&A..36..599B} among other historical references), radiative transfer has been included only recently in cosmological simulations of the formation of large scale structure in the Universe (see e.g. \citet{1999ApJ...523...66A, 2001NewA....6..437G, 2001MNRAS.324..381C, 2002ApJ...572..695R} and more recently \citet{2006MNRAS.369.1625I,2007ApJ...671....1T,2007MNRAS.377.1043M,2009A&A...495..389B}) . Among many different astrophysical problems that require a proper treatment of light propagation, cosmic reionization stands out as a particularly challenging one, because the ionizing radiation field plays a key role in the transition from the "dark ages" to the era of galaxy formation~: the chronometry and the geometry of the process is entirely related to the way matter and radiation interact. The proper numerical modeling of cosmic reionization represents a additional challenge, since it requires to capture a whole set of physical phenomena which are difficult to tackle on their own (see e.g. the review by \citet{2001PhR...349..125B}). In a nutshell, reionization can be described as "atoms being dissociated by UV photons emitted by stars formed in collapsed, self-gravitating halos".  This requires to follow the dynamics of dark matter and gas on large scale, cooling and star formation on galactic scales, the emission of ionizing radiation at microscopic scales and finally, UV light propagation back to the cosmological scales.  Because of this chain of causality involving many different cosmological fluids (dark matter, gas, stars and photons), it is only recently that significant progresses were made in the field of cosmological radiative transfer.

Computer simulations of radiative transfer cover a wide range of techniques, most of them reviewed in \citet{2009arXiv0906.4348T} and with most implementations gathered in two sets of comparison papers \citep{2006MNRAS.371.1057I,2009MNRAS.400.1283I}. Current cosmological radiative transfer codes successfully pass these rather academic tests, but it should be noted that only a few observational tests can be used as a probe to calibrate these rather complicate numerical tools. The first major constraint comes from quasars with the detection of Gunn-Peterson troughs and a decrease of the flux transmission in spectra of objects at z$\sim$6, which can be interpreted as the mark of the transition from a neutral Universe to an ionized one (see e.g. \citet{2004AJ....127.2598S}, \citet{2006AJ....132..117F}). From the observed spectra and provided that some assumptions are made on the structure of the density field or the UV background, important quantities such as mean free path, photoionization rate or UV field intensity can be constrained (see e.g. \citet{2002AJ....123.1247F}, \citet{2006AJ....132..117F}, \citet{2007MNRAS.382..325B}). These constraints provide anchor values at $z\sim 6$ for the calibration of cosmological simulations of reionization and track their ability to simulate the post-overlap era and the overlap itself (see e.g. \citet{2006ApJ...648....1G}). However, this technique only provide upper/lower boundaries at higher redshifts as complete absorption can be reached with a neutral fraction as low as 0.001. Furthermore, since models are used to infer physical properties from flux transmission, any agreement or disagreement between calculations and quantities derived from observations should be taken with caution (as noted by e.g. \citet{2007ApJ...671....1T}) and in a reversed role the simulations may happen to be informative about the proper way to interpret data. The second set of constraints comes from the scattering of CMB photons by electrons released during the reionization process. Usually expressed in terms of the Thomson optical depth $\tau$, current constraints from WMAP set $\tau=0.084\pm0.016$ implying a redshift of (instantaneous) reionization of $z\sim10.9 \pm 1.4$ \citet{2009ApJS..180..330K}. This constraint results from the integrated impact of the electrons on the CMB properties and is therefore more sensitive to the complete history of cosmic reionization.

In this paper, our goal is to confront our new radiative transfer code {\tt ATON} to these observational constraints, using a set of hydrodynamical simulations at different resolutions. This code has already been presented and tested using a standard test suite in \citet{2008MNRAS.387..295A}. The dynamical simulations include gravity and gas physics with mesh refinement, as well as widely adopted and well tested star formation recipe. The radiative transfer is performed as a post-processing step (full coupling of hydrodynamics with radiation is currently underway). It relies on a moment-based description of the propagation of light in the same spirit as e.g. \citet{2001NewA....6..437G} or \citet{2009MNRAS.393.1090F}. The original {\tt ATON} code has since been fully ported on Graphics Processing Units (GPU hereafter) architecture using CUDA. Thanks to the high acceleration rate ($\sim 100\times$ compared to CPU) made possible by such hardware, we have been able to simulate the radiative transfer at the same resolution as the hydrodynamics base grid with $1024^3$ cells.  The current article aims at reaching two objectives~: first, showing the ability of {\tt ATON} to model properly the reionization process and second, demonstrate the potential of GPU architecture for numerical cosmology. Regarding the ability to model the reionization, we partially recover the observational constraints at $z\sim 6$ if we include a simple clumping factor model. However we also find that the properties of the radiation field and the neutral fraction distribution are driven by very different regions, making it difficult to relate the average UV intensity to the average fraction of neutral gas. Regarding the adaptation of our code on GPU, we describe in details in the Appendix how such architecture can be used at full power on this type of problems.

This paper is organized as follows:~first we describe the methodology and the simulations. Second we describe a first set of fiducial simulations and assess in particular the issues related to resolution and numerical convergence. Third, we introduce a simple prescription for the subgrid clumping obtained from our most resolved simulation (12.5 Mpc/h with $1024^3$ dark matter particles) and apply it to the largest simulation we have (100 Mpc/h with $1024^3$ dark matter particles). Finally, we discuss our results, forthcoming applications and possible improvements.

\section{Methodology} 

\subsection{Simulations} 

The cosmological simulations analyzed in this work were produced using {\tt RAMSES} \citep{2002A&A...385..337T}. The cosmological parameters follow the WMAP-5 constraints \citep{2009ApJS..180..330K} and the initial conditions were generated using the MPGrafic package \citep{2008ApJS..178..179P}.
We have generated 4 sets of Gaussian random fields with different box sizes, based however on the same Poisson shot noise, so that the same structure should form at the same location, although with different timings. The number of cells and dark matter particles was set to $1024^3$ and we allow for 2 more levels of refinement, resulting in the mass and spatial resolution elements quoted in Table~\ref{t:boost}. The grid was dynamically refined up to the maximum allowed resolution, using a quasi-Lagrangian strategy: when the dark matter or baryons mass in a cell reaches 8 times the initial mass resolution, it is split into 8 children cells.

Gas dynamics is modeled using a second-order unsplit Godunov scheme \citep{2002A&A...385..337T, Teyssier:2006p413, Fromang:2006p400}
based on the HLLC Riemann solver \citep{Toro:1994p1151}. We assume a perfect gas Equation of State (EoS) with $\gamma=5/3$. Gas metallicity 
is advected as a passive scalar, and is self-consistently accounted for in the cooling function. Note that in the present work, no radiation background was considered for the cosmological simulation. As gas cools down and settles into centrifugally supported discs, we need to provide a realistic model for the interstellar medium
(ISM). Since the ISM is inherently multiphase and highly turbulent, it is beyond the scope of present-day cosmological simulations to try to simulate it self-consistently. It is customary to rely on subgrid models, providing an effective EOS that capture the basic turbulent
and thermal properties of this gas. Models with various degrees of complexity have been proposed in the literature \citep{Yepes:1997p1245, Springel:2003p1288, Schaye:2008p1393}. We follow the simple approach based on a temperature floor given by a polytropic EOS for gas 
\begin{equation}
T_{floor} = T_* \left( \frac{n_{\rm H}}{n_*} \right) ^{\Gamma -1}
\end{equation}
where $n_*=0.1$ H/cc is the density threshold that defines the star forming gas, $T_*=10^4$ K is a typical temperature mimicking
both thermal and turbulent motions in the ISM and $\Gamma=5/3$ is the polytropic index controlling the stiffness of the EOS. Gas is able to heat above this floor, but cannot cool down below it. Note that because of this temperature floor, the Jeans length in our galactic discs is always resolved. We also consider star formation using a similar phenomenological approach. In each cell with gas density larger than $n_*$, we spawn new star particles at a rate given by
\begin{equation}
\dot \rho_{*} = \epsilon_* \frac{\rho_{gas}}{t_{\rm ff}}~~~{\rm with}~~~t_{\rm ff}=\sqrt{\frac{3\pi}{32G\rho}}
\end{equation}
where $t_{\rm ff}$ is the free-fall time of the gaseous component and $\epsilon_*=0.01$ is the star formation efficiency. The star particle
mass depends on the resolution (see Table~\ref{t:boost}). For each star particle, we assume that 10\% of its mass will go supernova after 10 Myr. We consider a supernova energy of $10^{51}$ erg and one M$_\odot$ of ejected metals per 10 M$_\odot$ average progenitor mass. This supernovae feedback was implemented in the {\tt RAMSES} code using the "delayed cooling" scheme \citep{Stinson:2006p1402}. 

To summarize, we used for this simulation suite rather standard galaxy formation recipe, which have proven only recently to be quite successful in reproducing the properties of field spirals \citep{Mayer:2008p1478, Governato:2009p1455, Governato:2010p1442} and dwarf galaxies. The only missing ingredient is the radiation field, which will be considered in a second step using our radiation solver.

\subsection{Radiative Transfer} 

Each snapshot of the simulations is post-processed using the {\tt ATON} code, described and tested in details in \citet{2008MNRAS.387..295A} and briefly summarized in this section. The method relies on a momentum description of the radiative transfer equations with an M1 closure relation \citep{2007A&A...464..429G}. Radiation is described in terms of the first three moments of the distribution function of photons, the radiative energy density $N$ and the radiative flux $\bf F$ and the radiative pressure tensor $\bf P$. These quantities are averaged over a group of frequencies and satisfy the usual conservation relations: 
\begin{eqnarray}
	\frac{
	\partial N}{
	\partial t}+\bf{\nabla} {\bf F}&=&-\kappa N + S,\\
	\frac{
	\partial \bf F}{
	\partial t}+c^2\bf{\nabla} {\bf P}&=&-\kappa \bf F, 
\end{eqnarray}
where $\kappa$ stands for the local absorption rate and $S({\bf x,t})$ is the source field which includes the production sites of photons as well as the recombination radiation. The Eddington tensor $\bf D$ close the system through an equation of state: 
\begin{equation}
	{\bf P}= {\bf D} N, 
\end{equation}
where $\bf D$ is approximated by the M1 model \citep{Dubroca1999915} : 
\begin{equation}
	{\bf D}=\frac{3\chi-1}{2}{\bf I}+\frac{1-\chi}{2}{\bf n}\otimes{\bf n}. 
\end{equation}
The quantity $\chi$ depends only on the reduced radiation flux $f=|{\bf F}|/cN$ and spans the values from 1/3 (pure diffusion regime) to 1 (pure transport regime) and depends only on the local properties of the radiation fields. The exact formula for $\chi$ can be found in \cite{2008MNRAS.387..295A}. Such a formulation guarantees that the two extreme regimes are properly captured, while all intermediate situations are approximated by a superposition of diffusion and transport. It should also be noted that this scheme differs from the common first-order flux limiter approach by its ability to cast shadows behind absorbants \citep[see][]{2008MNRAS.387..295A}. 

The previous radiation conservation laws are solved using an explicit time integration, resulting in a stringent CFL conditions on the time step due to the high value of the speed of light~: 
\begin{equation}
	\frac{\Delta x}{c}>\Delta t.
\end{equation}
However, thanks to GPU acceleration, we can speed up each individual time step so that the resulting scheme can still achieve high performance. The details of the GPUs implementation are given in the Appendix. Originally the code is able to evaluate intercell fluxes using both the Haardt-Lax-van Leer (HLL) or the simpler Lax-Friedrich (LF) scheme but only the latter has been used in the current work. 

The photo-chemistry in {\tt ATON} is currently limited to Hydrogen with the associated cooling processes. Again, the energy conservation and the chemistry are solved in an explicit fashion and are sub-cycled during a radiative transfer step using a scheme in the spirit of \citet{1997NewA....2..209A}. It turns out that most of the time the characteristic time scales involved in cooling and chemistry are longer than radiative time steps, thus limiting the impact of `microphysics' calculations on the overall computation. 

All the processes (transport, cooling, and chemistry) and their equations are solved in a single frequency group where $\nu>13.6$ eV and involve average quantities such as the hydrogen photoionization cross-sections $\sigma_e=2.49\cdot 10^{-18}\mathrm{cm}^2$ (energy averaged), $\sigma_n=2.93 \cdot10^{-18} \mathrm{cm}^2$ (number averaged) and the typical ionizing photon energy $e=20.27$ eV, where a 50 000K black body spectrum is assumed.

Typically, one complete radiative transfer simulation requires between 30 000 and 240 000 time steps depending on the resolution which defines the time step and the starting redshift. The 800 Myr of cosmic evolution we would like to cover is described with a time resolution of 3500 years. The code has been deployed on GPU architecture using the API CUDA 2.2, (becoming thus {\tt CUDATON}) developed for devices built by the Nvidia company. The code runs independently from the CPU, without any transfer between the host and the device except during the initial setup and for the outputs on hard drives. The typical acceleration observed compared to single-cpu runs is close to a 100. Using an additional MPI layer, {\tt CUDATON} is able to run on multi-gpu architecture with communications between the devices, which requires additional transfer between hosts and GPUs. The additional cost is close to 10 percent of the total computing time since data have to be transfered through PCI-Express ports. All the calculations here were performed on 128 Tesla C1060 devices on the Titane supercomputer of the CCRT computing center. Typically a single radiative post-processing run on a $1024^3$ grid is performed in 2.5 hours but can be as short a 1 hour for coarse simulations with simple physics and as long as 18 hours for our most realistic calculations. During the course of this project, a couple hundred of calculations over six months were performed to improve the code and to test our various recipes. 

\subsection{Source modeling for the radiative post-processing}

The sources of photons, namely young stars, are produced by the cosmological simulations that return for each stellar particle its position, velocity, age, mass, and metallicity. From there, the source modeling is inspired from the procedure described in \citet{2009A&A...495..389B}. Stellar particles are assumed to satisfy a Salpeter IMF resulting in a global spectra well approximated by a 50 000 K black body. Individual lifetimes of stellar particles as ionizing sources are drawn randomly between 5 and 20 Myrs. Overall, for each source, the production of ionizing photons lies between 24 000 and 98 000 per stellar baryon over its lifetime. Because the sources appear at discrete times, due to the discontinuous production of snapshots, the sources contribution is smoothed out over all the duration between 2 successive snapshots using the following strategy. When modeling the radiative transfer from time $t_p$ to $t_{p+1}$, we consider only star particles contained in snapshot $p+1$. Knowing their age $a_{p+1}$, we calculate their age $a_p=a_{p+1}-(t_{p+1}-t_{p})$ at time step $p$, which can be negative if the star appeared at a time $a^*$ between the two snapshots. Then~: 
\begin{itemize}
	\item if $a_p$ is greater than the source's lifetime $L$, it is discarded, 
	\item if $a_p<0$, the source has been created between the two snapshots. However, it will contribute to the photon emission from $t_p$ to $t_{p+1}$ with a `diluted photon' emission rate given by $(a_{p+1}-a^*)/(t_{p+1}-t_{p})\dot N$, 
	\item if $0<a_p<L$ the source will have an emission rate given by $\mathrm{min}(1,L-a_p)/(t_{p+1}-t_{p})\dot N$. If the source ends its ionizing phase between the snapshots, it will nevertheless contribute continuously from $t_p$ to $t_{p+1}$ with a `diluted' emission rate. 
\end{itemize}
From there, sources are projected on the 3D grid using the Nearest Grid Point assignment scheme.  Emission is modeled as a field where each cell acts as a single photon source.

These intrinsic luminosities are modulated by two additional factors to give the effective source luminosity: 
\begin{equation}
	\dot N_\mathrm{eff}=\dot N\times f_\mathrm{esc} \times B. 
\end{equation}
The first factor is the escape fraction $f_\mathrm{esc}$ which models the actual fraction of radiative energy which manage to escape the stellar environment. Typical values can be as high as $20\%$ and is essentially a free parameter which allows to tune the reionization redshift. The second factor, $B$, is called here the \textit{boost factor}. It is a correction term that compensates from the unresolved star forming halos in the simulation, a major resolution effect on the simulated star formation history. As shown in e.g. \citet{2006A&A...445....1R}, the mass resolution has a significant impact on the star formation history (SFH) if large simulation volumes are considered, when the minimum resolved halo mass (optimistically set to 100 dark matter particles) is larger than the minimum mass for star forming halos, based on atomic cooling arguments \citep{Gnedin:2000p1555, 2006A&A...445....1R, Hoeft:2006p1565}. This minimum mass (also referred to as the Filtering Mass) starts around $10^7$ M$_\odot$ before reionization and then rises steadily as $(1+z)^{3/2}$ from redshift 6 to the final epoch \citep{2006A&A...445....1R, Hoeft:2006p1565}. Resolving this minimum mass before reionization will require a dark matter particle mass below $10^5$~M$_\odot$, a rather strong requirement for cosmological simulation. Only our smallest box size (12.5 Mpc/h with 1024$^3$ particles) barely satisfies this criterion.  

\begin{figure}
	\center{
	\includegraphics[width=\columnwidth]{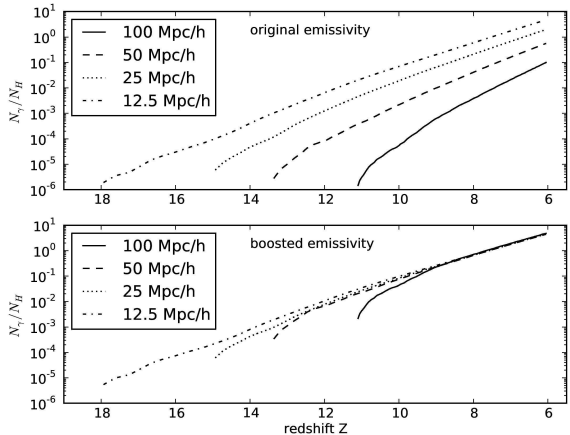}} 
	\caption{Integrated number of emitted photons in units of the number of hydrogen atoms as a function of redshift in four ($1024^3$+2 levels of refinement) hydrodynamical simulations with different coarse resolutions. The top panel shows the original emissivity due to the simulated stellar population, while the bottom panel shows the boosted emissivity which compensates the impact of resolution on the simulated SFR. All plots are performed with $f_\mathrm{esc}=0.03$.} 
	\label{f:nlum} 
\end{figure}

As an illustration, the top panel of Figure~\ref{f:nlum} shows the evolution of the integrated number of photons with time in 4 simulations at different resolutions, which depends directly on the simulated SFR. Clearly the difference in resolution has an impact on the apparition of the first sources~: low resolution simulations require a longer time to reach the epoch of the formation of the first stars~: $z\sim 11$ for the 100 Mpc/h simulation versus $z\sim 18$ for the 12.5 Mpc/h simulation. Furthermore, this late start is not compensated by a higher SFR and at $z=6$, the number of emitted photons decreases as lower spatial resolution are considered. 

We used the analytical model of  \citet{2006A&A...445....1R} to compute the expected converged star formation history. We can compensate for the unresolved star forming halos by boosting each resolved UV emitting sources by a "boost factor", derived to put the actual simulated SFRs (and hence the number of emitted photons) in accordance with the converged one. We have used for the boost factor the following simple functional form: 
\begin{equation}
	B(t)=\mathrm{min}(1,a_b\exp(k_b/t))=\frac{\mathrm{SFR}_\mathrm{converged}(t)}{\mathrm{SFR}_\mathrm{actual}(t)}, 
\end{equation}
where $t$ is the age of the Universe. The parameters $a_b$ and $k_b$ are fitted in the measured SFH in each simulation. They hence depend on the resolution and are given in Table \ref{t:boost} for $1024^3$ + 2 levels of refinement simulations with the WMAP-5 cosmology. The resulting integrated photon numbers is shown in the bottom panel of Figure \ref{f:nlum} and exhibit a good level of convergence at redshifts $z<9$. Let us emphasize that the two parameters, $f_\mathrm{esc}$ and $B$ are different by nature~: $B$ is not a free parameter and follow from the proper analysis of resolution effect and is in some sense a pure numerical correction. Meanwhile, $f_\mathrm{esc}$ remains as a physical parameter which models e.g. the subgrid physics and in the end, serves mostly to set the redshift of reionization. Of course, this simple prescription does not fix the late apparition of stars at low resolution since it only corrects existing stars without creating new sites of stellar formation. In our investigations, it appears that low resolution simulations (with the largest correction) exhibit similar behaviors than highly resolved ones but we admittedly focus on average quantities and global distributions. Fine geometrical details, on the other hand, are likely to be poorly captured by this boost factor approach, because of the lack of small emitters in unresolved site of stellar formation. 

\section{Results} 

Several aspects were investigated during this work, starting from basic numerical experiments focusing mostly on the resolution effects to slightly more complex modelisation where we attempt to fit observational data. First, we describe our fiducial results obtained from the post-processing of AMR simulations. From there, we discuss the impact of sub-cell clustering to the modelisation with a focus on the quantity of absorbants at $z\sim 6$.

\subsection{Fiducial experiments}

\subsubsection{Global properties} 

The fiducial experiments consist in four simulations described in Table \ref{t:boost}, with comoving box size ranging from 100 Mpc/h to 12.5 Mpc/h. The dynamical simulations were performed on a $1024^3$ coarse grid + 2 level of refinement at $z \sim  6$ while the radiative transfer post-processing was computed on a $1024^3$ regular grid. Highly resolved simulations are expected to better resolve the small scales photon sinks but lack the strong and rare sources that populate large scale volumes. Conversely, large simulation have a better representativity of rare and strong events but lack the resolution on absorbants. This features are somehow reflected in the values of the escape fraction shown in Table \ref{t:boost}~: for a redshift of reionization chosen to be $z_\mathrm{ion}=6.5$, $f_\mathrm{esc}$ decreases with the box size from 0.055 to 0.02. Highly resolved simulations have sources embedded in highly clustered gas, implying a more efficient recombination, and these sources cannot be as strong as the ones found in large volumes. Overall, such simulations require a larger amount of photons to reionize. 

\begin{figure}
\center{
\includegraphics[width=\columnwidth]{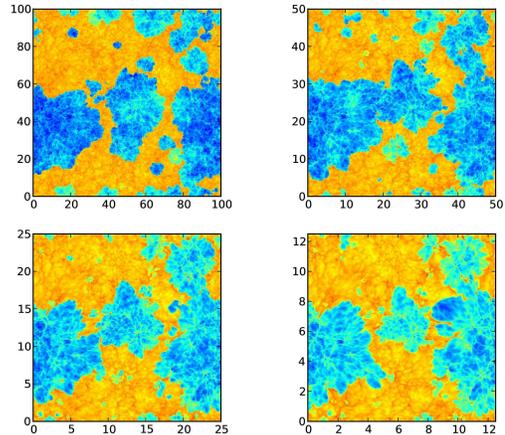}} 
\caption{Neutral hydrogen density maps at $z\sim 7.3$ and $x\sim 0.5$ (volume weighted) for boxes of comoving length 100, 50, 25, 12.5 Mpc/h. All maps have a resolution of $1024^2$ and a thickness of 5 Mpc/h. The color scale is logarithmic with blue and red regions standing respectively for low and high density of neutral hydrogen. Coordinates are expressed in comoving Mpc/h.} 
\label{f:lmap} 
\end{figure}

Maps of the distribution of neutral gas are shown in Fig. \ref{f:lmap} at half reionization. Let us recall that these four simulations were performed with initial conditions that shared the same set of phases leading to similarities in the global spatial distributions. These maps exhibit the expected global behaviors: high resolution simulations present complex ionization fronts, which result from the highly inhomogeneous structure of the absorbing regions. Meanwhile, low resolution simulations fail to resolve small scale structures leading to smoother fronts. Looking at the details of zoomed maps (see the bottom plots in Fig.\ref{f:lmap}), highly resolved simulations present dense neutral clumps within ionized regions whereas these absorbants are absent from large under-resolved boxes. The failure of large simulations boxes to resolve these small scales will prove to be crucial in our ability to reproduce the data at $z\sim 6$. 


\begin{table*}
\begin{center}
\begin{tabular}{|c|c|c|c|c|c|c|}
\hline
Box size & $a_b$ & $k_b$ & $f_\mathrm{esc}$ & $m_\mathrm{dm}$ & $m_\mathrm{bar}$ &$m_\mathrm{star}$ \\
Mpc/h &    & Myr &  & $M_\odot$ & $M_\odot$ &$M_\odot$ \\
\hline
12.5 & 0.7 & 300 & 0.055 &1.52$\times10^5$&2.54$\times10^4$ & 5.81 $\times 10^4$\\
25 & 1.0 & 650 & 0.030 &1.22$\times10^6$&2.03$\times10^5$&4.65 $\times 10^5$\\
50 & 1.2 & 1500 & 0.020&9.76$\times10^6$&1.62$\times 10^6$&3.72 $\times 10^6$\\
100 & 1.2 & 3000 & 0.020&7.81$\times10^7$&1.30$\times 10^7$ &2.97$\times 10^7$\\
\hline
\end{tabular}
\end{center}
\caption{Summary of the parameters used in our simulation suite. Parameters $a_b$ and $k_b$ are used in the SFR correction to account for finite resolution, assuming WMAP-5 cosmology. $f_\mathrm{esc}$ is the assumed escape fraction, $m_\mathrm{dm}$ is the mass resolution of dark matter particles while $m_\mathrm{bar}$  is the mass resolution per AMR grid cell. Also shown is the minimum star particle mass. All simulations were performed with $1024^3$ dark matter particles.} 
\label{t:boost} 
\end{table*}

\begin{figure}
\center{
\includegraphics[width=\columnwidth]{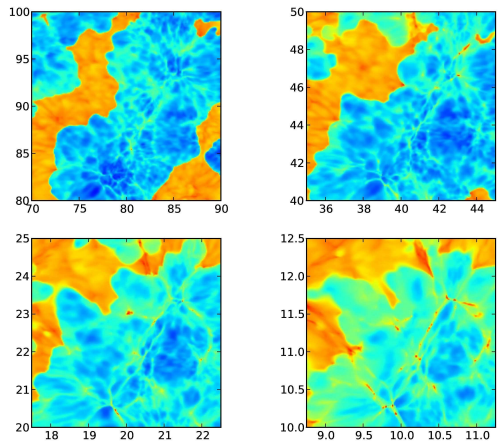}} 
\caption{Same as Figure~\ref{f:lmap} but zooming on a photoionized region. Coordinates are expressed in comoving Mpc/h.} 
\label{f:lmapzoom} 
\end{figure}

\subsubsection{Neutral fraction} 

\begin{figure}
\center{
\includegraphics[width=\columnwidth]{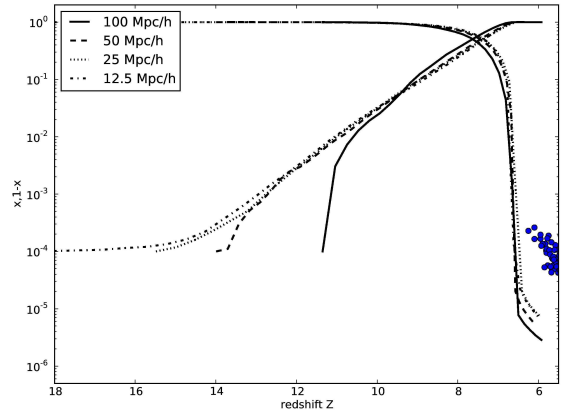}} 
\caption{Neutral and ionised volume-averaged fraction as a function of redshift measured in the 100, 50, 25 and 12.5 Mpc/h boxes. Dots stand for observationnal constraints given by \citet{2006AJ....132..117F}.} 
\label{f:xv_avg} 
\end{figure}

\begin{figure}
\center{
\includegraphics[width=\columnwidth]{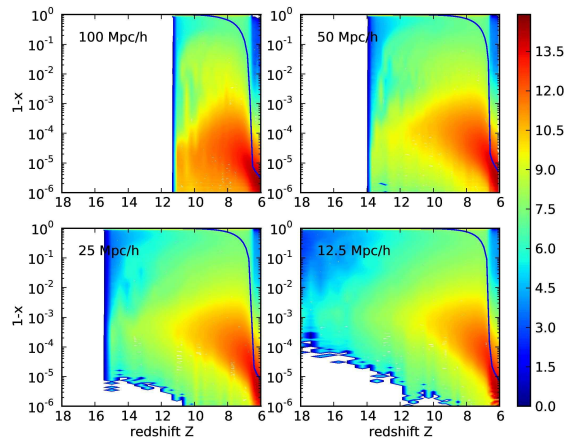}} 
\caption{The evolution of the hydrogen neutral fraction $f_{HI}$ in the 100, 50, 25 and 12.5 Mpc/h boxes (\emph{from top to bottom}). The blue lines stand for the evolution of the volume-weighted average value and the color levels show the evolution of the $f_{HI}$ distribution with redshift. Values with a high probability are shown in red and values with a low probability are shown in blue.}
\label{f:xv_dist} 
\end{figure}

The evolution of the volume-weighted ionized fraction $x_v$ and neutral fraction $1-x_v$ is shown in Fig. \ref{f:xv_avg}. Escape fractions were chosen to achieve reionization at $z\sim 6.5$ and it can be seen that all four calculations present similar behavior for $z<9$. Distribution of values are shown as colored contours in Fig. \ref{f:xv_dist} and it can be seen that $x_v$ is representative of the distribution of ionized fraction in the boxes as it tracks accurately the most probable value. For earlier times, notable differences arise from the impact of resolution on star formation and the production of photons. Highly resolved simulations have ionization history that expand up to $z=18$ where their first stars form. Conversely the largest simulation forms stars only at $z=11$. Because of the introduction of a time-dependent boost of their sources luminosity, these large simulations quickly catch up the highly resolved one, resulting in a strong slope for $x_v$. The `catching up' effect is clearly noticeable in the 100 Mpc/h calculation but already much limited for the 50 Mpc/h simulation, almost unnoticeable for the 25 Mpc/h box and for $z>9$ the calculations have all converged. This should not come as a surprise since the boost factor approach was designed precisely for that reason. Nevertheless, the different clustering of gas, of sources as well as their number could have resulted in a difference in the ionization history of the different calculations even though they share the same global amount of photons emitted. Since we do not observe such a discrepancy, it suggests that small photon sources missing from the large boxes are located roughly at the location of the large photon sources present in these boxes: by boosting their luminosity, we compensate at the sub-grid level for the lack of stellar particles at the correct location. 

Let us also point out that the neutral fraction calculated at $z=6$ spans from $3\times10^{-6}$ for the 100 Mpc/h simulation to $10^{-5}$ for the 12.5 Mpc/h. Such levels of neutral fraction are inconsistent with constraints provided by \citet{2006AJ....132..117F} from Gunn-Peterson troughs in quasars spectra which imply a typical level of $10^{-4}$ at z=6. Even though this estimation relies on assumption on the distribution of gas and on an homogeneous field of radiation, this level of neutral gas has been reproduced by e.g. \citet{2007ApJ...671....1T}, \citet{2007ApJ...657...15K}, \citet{2008ApJ...681..756S} and on highly resolved simulations by \citet{2006ApJ...648....1G}. On the other hand, \citet{2009MNRAS.400.1049F} present the same level of discrepancy at admittedly much lower resolution. We investigate this point further on in subsequent sections, but at the current stage, all our simulations fail to reproduce the observed neutral fraction without additional modelisation.

\subsubsection{Radiation field-UV intensity} 

\begin{figure}
\center{
\includegraphics[width=\columnwidth]{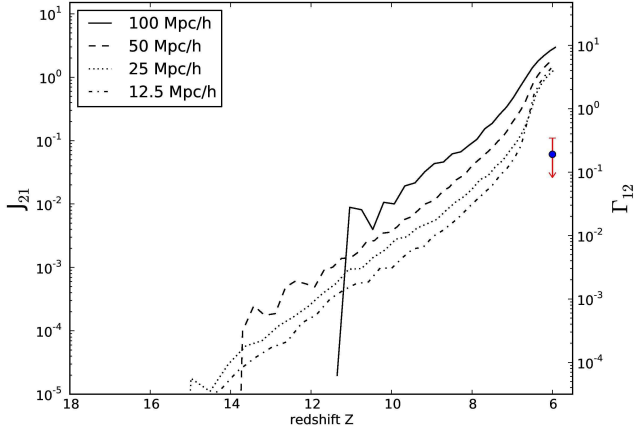}} 
\caption{Evolution of the mean intensity of ionizing radiation in the 100, 50, 25 and 12.5 Mpc/h boxes. The upper limit at $z\sim 6$ stands for the constraint given by \citet{2007MNRAS.382..325B}.} 
\label{f:J21avg} 
\end{figure}

The moment-based description of radiative transfer allows us to track the radiation intensity in each cell, usually described in terms of $J_{21}$ or in the terms of photoionization rate $\Gamma_{12}$. The evolution of the volume averaged intensity is shown in Figure \ref{f:J21avg} as well as the constraint provided by \citet{2007MNRAS.382..325B}. For our four boxes, the evolution of the radiation intensity exhibits a `cobra-like' shape with a sharp increase until the reionization epoch, during which the increase is the steepest, followed immediately after by a flattening of the slope at the end of reionization. The amount of radiation is larger by a factor of 3 at z=6 when comparing the 100 Mpc/h and the 12.5 Mpc/h box, while the 25 Mpc/h and 12.5 Mpc/h calculations seem to have converged. This trend is consistent with the differences in the residual neutral fraction calculated at z=6 (see Figure \ref{f:xv_avg}), where the highly resolved calculations present a larger amount of neutral gas than the poorly resolved ones. A stronger radiation field imply a larger photoionization rate and therefore a smaller amount of neutrals. 

\begin{figure}
\center{
\includegraphics[width=\columnwidth]{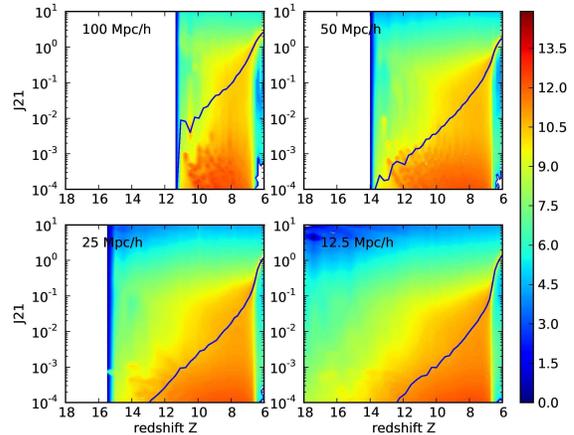}} 
\caption{Evolution of the mean intensity $J_{21}$ in the 100, 50, 25 and 12.5 Mpc/h boxes (\emph{from top to bottom}). The blue lines stand for the evolution of the volume-weighted average value and the color levels show the evolution of the $J_{21}$ distribution with redshift. Values with a high probability are shown in red and values with a low probability are shown in blue.} 
\label{f:j21_dist} 
\end{figure}

Unfortunately, these calculations all agree on one point: they are inconsistent with observational constraints, such as the one provided by \citet{2007MNRAS.382..325B} by a factor of 20 to 50. Again, such a discrepancy has recently been found at lower resolution by \citet{2009MNRAS.400.1049F} using an alternative implementation of moment based radiative transfer. This excess of radiation goes along with the lack of neutral gas at the end of reionization in our computations. Furthermore, an inspection of Figure \ref{f:j21_dist} reveals that the average intensity is representative of the most probable value in the boxes, discarding any possibility of a biased value.

\subsubsection{Optical Depth} 

\begin{figure}
\center{
\includegraphics[width=0.8\columnwidth]{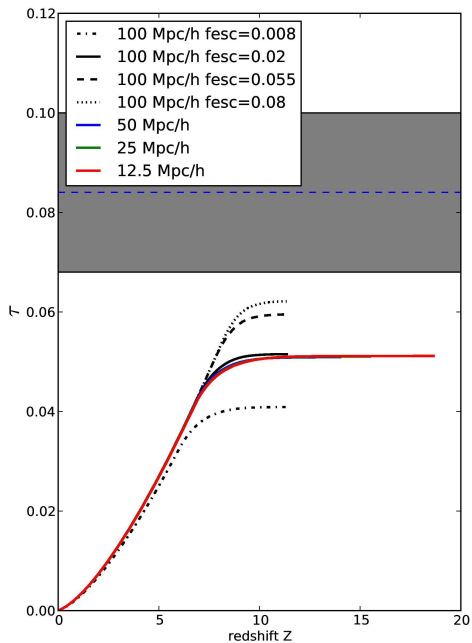}} 
\caption{The Thomson optical depth computed from the average mass weighted
  electron density. The gray area stands for the WMAP-5 measurements allowed
  range at $1\sigma$ level (see \citet{2009ApJS..180..330K}). The 50, 25 and
  12.5 Mpc/h curves are almost superimposed.} 
\label{f:tau} 
\end{figure}

The Thomson scattering of CMB photons by the electrons released during the reionization is quantified through the Thomson Optical depth given by 
\begin{equation}
\tau=c\sigma_t \int_{z_\mathrm{rec}}^0 n_e(z) \frac{d t}{d z} dz, 
\end{equation}
where $\sigma_t$ is the corresponding cross-section and $n_e=x n_H$ is the density of electrons released by ionized hydrogen atoms. Our calculations of the optical depth is presented in Figure \ref{f:tau} for the four simulations. Also presented is the constraints range obtained from the 5 years release of CMB measurements made by the WMAP collaboration \citep{2009ApJS..180..330K} at the $1\sigma$ level.

\begin{figure}
\center{
\includegraphics[width=\columnwidth]{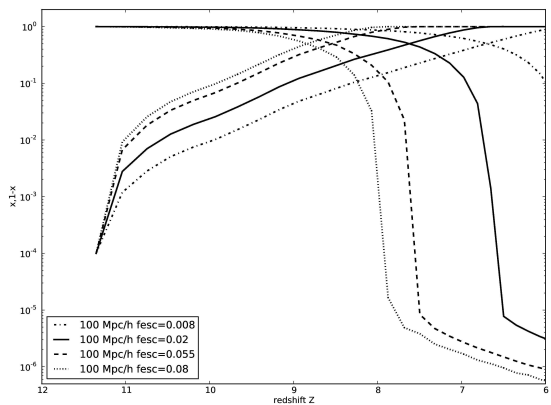}} \caption{Evolution of the volume weighted neutral and ionized hydrogen fraction in comoving 100 Mpc/h-$1024^3$ boxes with different escape fractions.} \label{f:fescvar} 
\end{figure}

The four fiducial experiments were performed at different resolution and therefore exhibit different ionization histories but have converged in terms of optical depth. This agreement result from the fact that the bulk of electrons production lies within the convergence redshift range (z$<$11). The four of them present an optical depth $\tau=0.051$ which lies at $2\sigma$ from the CMB value. The inclusion of helium electrons would slightly increase the amount of electrons \citep[see e.g.][]{2009MNRAS.400.1049F} but probably not at levels that would make it consistent with the WMAP expected value. This discrepancy has already been noted by \citet{2006ApJ...648....1G}, \citet{2007ApJ...671....1T} or \citet{2009MNRAS.400.1049F} for simulations with similar ionization redshifts. Also shown in Figure \ref{f:tau} is the optical depth measured in the largest box (100 Mpc/h), but with larger and larger escape fractions. These calculations were performed at the same level of resolution than the fiducial experiments and under the same protocol. Clearly, the resulting $\tau$ gets closer to the CMB value as a consequence of a larger density of photons at earlier times. It indicates that a larger escape fraction could be the solution toward an agreement, however it comes at the cost of a larger redshift of reionization as shown in Figure \ref{f:fescvar} which would place the z=6 neutral fraction even further to the observed constraints than the fiducial ones. It is therefore likely that a plausible path toward an agreement between the observed and the calculated $\tau$ lies in a varying escape fraction which would ensure an extended ionization history to increase $\tau$ while keeping the reionization redshift reasonably low. For instance investigations by \citet{2009ApJ...693..984W} suggest that higher escape fractions could exist in small galaxies. Furthermore \citet{2007MNRAS.376..534I} argue that early populations in small haloes  at $z\sim 22$  are important contributors to the optical depth, and these objects are missing in our calculations. Other routes toward agreement may lie in additional physics such as the inclusion of specific population III sources \citep{2007ApJ...671....1T}. Finally it should be noted that \citet{2008ApJ...685....1S} suggest that inaccuracies in the reionization history and degeneracies in cosmological parameters lead to a larger range of possible values of $\tau$, 0.06-0.11 at $1\sigma$~: the discrepancy between our calculation and the CMB constraint could be resolved in between at $\tau\sim0.07$.

\subsubsection{Density dependance of UV intensity and neutral fraction}
 \label{s:densdep} 
 
\begin{figure*}
\center{
\includegraphics[width=1.5\columnwidth]{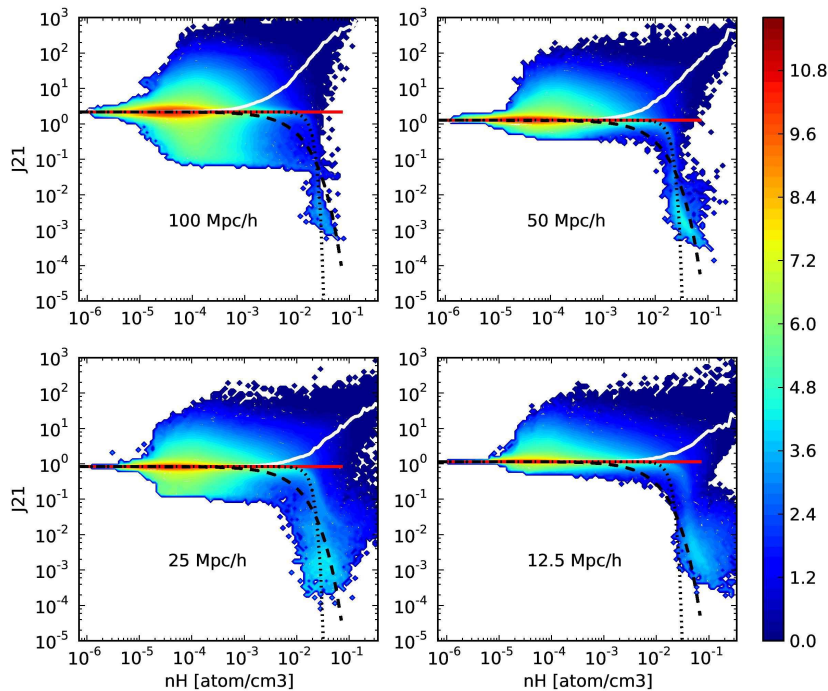}} 
\caption{The density contrast vs ionizing intensity (as $J_{21}$) relations measured in the 100, 50, 25 and 12.5 Mpc/h boxes at $z\sim6.25$, which corresponds to a fully ionized simulation. Red (resp. blue) regions stand for high (resp. low) probabilities in the distributions. The white line stand for the average ionizing intensity per density bin $\langle J_{21}\rangle (nH)$. The dashed red line stands for the volume average $J_{21}=J_0$ value, the dashed black line stand for the $J_{21}=J_0\exp(-n_H/n_1^*)$ model and the dotted line stand for the $J_{21}=J_0\exp(-(n_H/n_4^*)^4)$ model. } 
\label{f:J21nh} 
\end{figure*}

\begin{figure*}
\center{
\includegraphics[width=1.5\columnwidth]{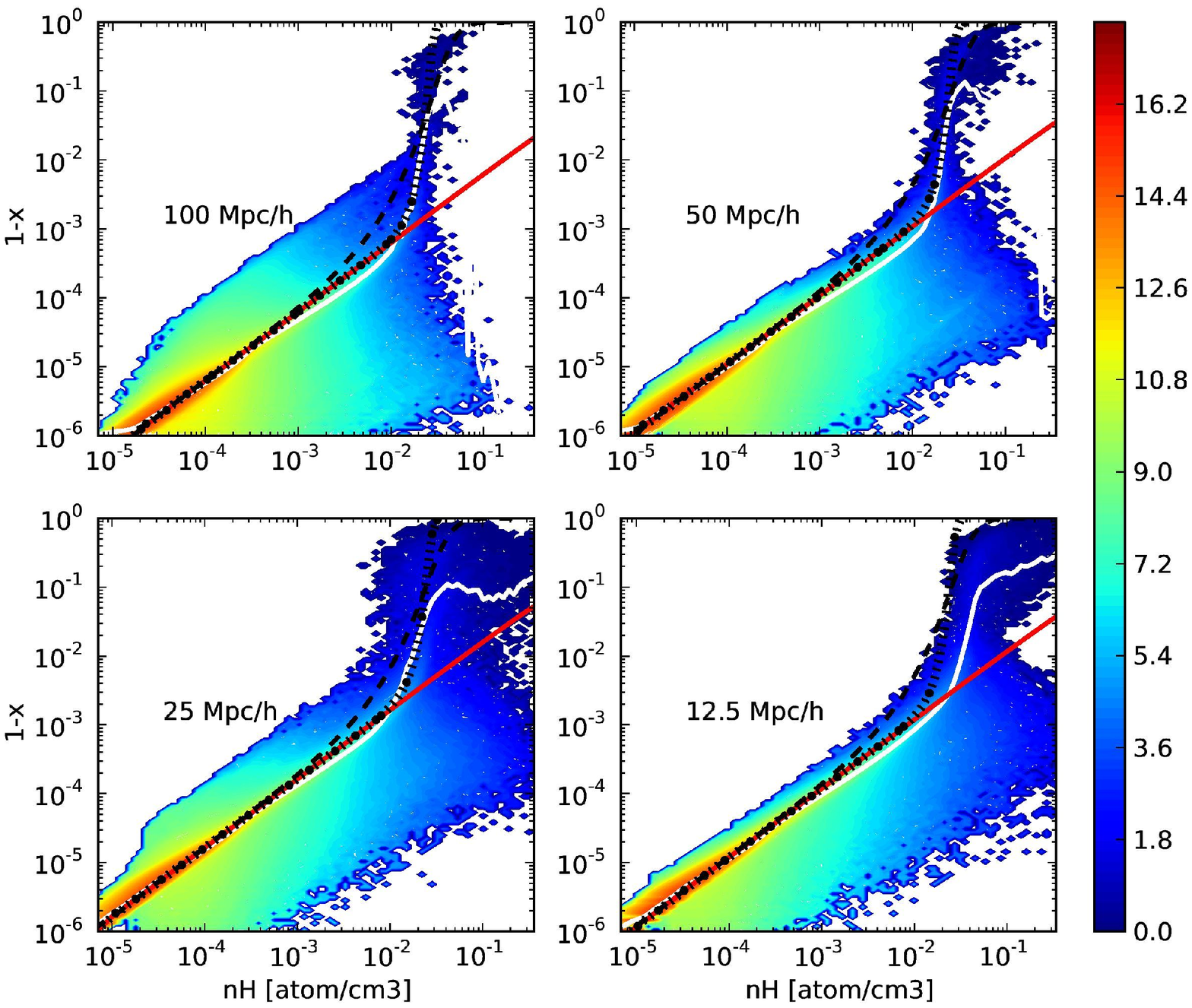}} 
\caption{The density contrast vs neutral fraction relations measured in the 100, 50, 25 and 12.5 Mpc/h boxes at z=6.25, which corresponds to a fully ionized simulation. Red (resp. blue) regions stand for high (resp. low) probabilities in the distributions. The white line stands for the average neutral fraction per density bin $\langle 1-x\rangle (n_H)$. The red line stand for the expected neutral fraction assuming equilibrium and an ionizing intensity equal to $J_0$. The dashed and dotted line stand respectively for the expected neutral fraction assuming equilibrium for the $J_{21}=J_0\exp(-n_H/n_1^*)$ and $J_{21}=J_0\exp(-(n_H/n_4^*)^4)$ models. } 
\label{f:xnh} 
\end{figure*}

In order to investigate our results on the {\it average} neutral fraction and UV intensity, we have computed the distribution $(1-x)(n_H)$ and $J_{21}(n_H)$ just after the reionization at z=6.25. The results are shown in Figures \ref{f:J21nh} and \ref{f:xnh}. From Figure \ref{f:J21nh}, we can see that the radiation field is not strictly homogeneous: although it is quasi-constant for densities $n_H<0.001$ $\mathrm{cm}^{-3}$, we see a significant decrease of the flux in the densest regions. In particular, an accumulation of obscured regions is apparent at densities close to $5\times10^{-2}$ cm$^{-3}$ with a radiation field 1000 times weaker than the volume average. This feature is more proeminent in highly resolved simulations (12.5 and 25 Mpc/h) and corresponds to a better treatment of dense, small absorbants, which we fail to model properly in large boxes. The strong cutoff of radiation in high-density regions is a manifestation of self-shielding, where dense clumps are protected from the ionizing background by their own high densities. 

We model this high-density behavior using two different models with different levels of exponential cutoff~: 
\begin{equation}
J_{1}(n_H)=J_0\exp(-n_H/n_1^*) 
\end{equation}
and 
\begin{equation}
J_{4}(n_H)=J_0\exp((-n_H/n_4^*)^4), 
\end{equation}
where $J_0$ stands for the average intensity field at low density and $n_{1,4}^*$ is the characteristic density at which the exponential cutoff operates. For sake of simplicity, we arbitrarily assigned the same values for the characteristic self-shielding densities for the four simulations, namely $n_{1}^*=0.007$ cm$^{-3}$ and $n_{4}^*=0.018$ cm$^{-3}$, which reproduce accurately the global $J_{21}(n_H)$ behaviors in the three largest boxes and is slightly off for the $12.5$ Mpc/h calculation. On the other hand, the plateau value is computed separately for the 4 box sizes. Clearly, the $J_4$ model is a better representation of the density dependence but the other model $J_1$ has also been considered for sake of comparison. It should be noted that the volume-average value $\langle J_{21}\rangle (n_H)$ per density bin is also shown on Figure \ref{f:J21nh}, as the white solid line. Surprisingly, it increases to levels up to $100$ times the volume average value, instead of following the exponential cutoff. While the latter is a good model for the {most probable} radiation flux, it does not predict the correct volume-average value. The discrepancy starts at densities close to $10^{-3}$ cm$^{-3}$ and is likely to be due to strong sources of radiation which are found inside galaxies. This illustrates the fact  that the radiative intensity may be subject to strong biasing effects, especially at high densities, and its average value must be therefore taken with caution. 

These models for the UV radiation field can be used to compute the expected neutral fraction, assuming photoionization equilibrium. The result of such a procedure is shown in Figure \ref{f:xnh} for the four fiducial simulations with the equilibrium $(1-x)(n_H)$ curves computed for a uniform radiation field equal to $J_0$ and for the small/strong cutoffs models $J_1$ and $J_4$. Also shown is the average neutral fraction per density bin. Clearly the average neutral fraction follows the equilibrium trend at low densities ($n_H<0.001$ cm$^{-3}$) which is well reproduced by the three models. For higher densities ($n_H>0.01$ cm$^{-3}$), the average neutral fraction rises sharply and its distribution presents a significant tail toward neutral gas, even though the spread remains quite important : for instance at $n_H\sim 2\times 10^{-2}$ neutral fractions from $10^{-6}$ to $1$ can be found with almost equal probabilities. This tail cannot be reproduced by the uniform UV field model $J_0$ which is not surprising since $J_0$ is not representative of the radiation field in which self-shielded high density regions lie. The small cutoff model $J_1$ does a better job at reproducing the tail but underestimates the strength of rise toward larger neutral fractions. The strong cutoff model $J_4$ is in better agreement which is also expected since it models more accurately the typical trend of the UV field as a function of density (see Figure \ref{f:J21nh}). This overall agreement between the computed radiation field and neutral fractions indicates that the global neutral fraction can be recovered assuming photoionization equilibrium, {\it as long as the correct model for self-shielded UV radiation is used}\footnote{Incidentally, it also shows that some consistency is achieved within our code}.

\subsection{Subgrid clumping model}

Our fiducial numerical experiments, albeit highly resolved in terms of radiative transfer, lack some resolution for the underlying gas distribution. From Table \ref{t:boost} the largest simulation fail to resolve star forming haloes at $z>11$ and $\sim 10^7 M_\odot$ mini-haloes which are expected to act as a sink of photons during the reionization process. On the other end of our sample of simulation, the smallest boxes reasonably resolve these scales but are too small to e.g. provide a correct description of the cosmological HII regions which are expected to be as large as tens of Mpc \citep[see e.g.][]{2001PhR...349..125B, 2007MNRAS.377.1043M}. From now on, we focus on the largest simulation (100 Mpc/h with $1024^3$ dark matter particles) but with an additional subgrid model, in order to combine large scale statistics with a corrected small-scale physical model. From now on, we only compare our calculations to the constraints at $z=6$ from quasars spectra, and put Thomson optical depth measurements aside. Since this quantity is more sensitive to the global reionization history, the fact that our star formation history starts at $z=11$ in the largest box cannot be compensated by any other means than just increasing the mass resolution or drastically changing the star formation recipe. Exploring these possibilities is postponed to future work.

\subsubsection{Clumping factors} 

When considering the hydrogen chemical balance equation, one gets: 
\begin{equation}
\frac{dn_{H_I}}{dt}=\alpha n_e n_{H_{II}} -\Gamma n_{H_I} 
\end{equation}
which is modified in the following manner if one consider the ionization fraction $x$~: 
\begin{equation}
-n_H\frac{dx}{dt}=\alpha n_H^2 x^2 -(1-x)n_H\Gamma, 
\end{equation}
where $n_H$ stands for the hydrogen number density (neutral+ionized), $\alpha$ and $\Gamma$ are respectively the recombination and photoionization coefficients, and $x$ is the usual ionized fraction. As we deal with fields defined on a grid, we only have access to quantities averaged within the cells such as $\langle n_H\rangle$ which lack some information on the subgrid variations. Defining a recombination clumping factor as $C_R=\langle n_H^2 x^2\rangle/\langle n_H x\rangle^2$ and a photon-atomic density cross clumping factor $C_I=\langle n_\gamma (1-x)n_h\rangle/\langle n_\gamma\rangle\langle (1-x) n_H\rangle$, the chemistry equation can be rewritten as: 
\begin{equation}
\frac{dx}{dt}=-(\alpha) \langle n_H\rangle ^2 x^2 C_R -(1-x) c \sigma \langle n_H \rangle \langle n_\gamma\rangle C_I. 
\end{equation}
The choice of definition for clumping factor is not unique, for instance \citet{2007ApJ...657...15K} define clumping factors where the averages are take over the whole terms such as $\langle \alpha(T) n_H^2 x^2\rangle$, which depends on density, ionization fraction and also temperature. Our choice of clumping factors basically assumes that temperature is distributed uniformly within the computational cells.

We compute the clumping factors for the 100 Mpc/h simulation using the 12.5 Mpc/h simulation. For $6<z<18$, the total Jeans Mass in the case of an adiabatically cooling gas decreases from $1.5\times 10^4M_\odot$ to $4\times 10^3M_\odot$ while the baryonic filtering mass evolves from $10^5 M_\odot$ to $5\times 10^4 M_\odot$ during the same intervall \citep[see e.g.][]{1998MNRAS.296...44G, 2001PhR...349..125B, 2007MNRAS.377.1043M}. From Table \ref{t:boost}, one can see that our mostly resolved simulation almost achieves this level of resolution. Our model should therefore provide a reasonable description of the density distribution at small scale.

\begin{figure}
\center{
\includegraphics[width=0.8\columnwidth]{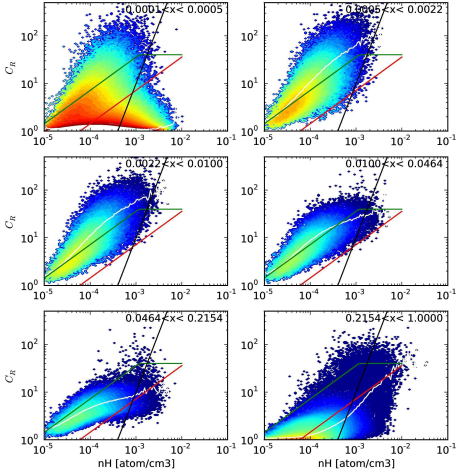}} 
\caption{Clumping factors $C_R$ as a function of the density computed from $8\times8\times8$ cells in the 12.5 Mpc/h simulation and used in the 100 Mpc/h experiment. The six panels show the $C_R$ distributions at different ionization levels~: the solid line shows the $\langle C_R \rangle (n_H)$ trend while the green dotted and the red dashed line stand for $C_R \sim n_H^{0.7}$ models with respectively a high and low normalization. The black dashed line stand for models with $C_R\sim n_H^{2.5}$ at high densities. } 
\label{f:clumpingR} 
\end{figure}

\begin{figure}
\center{
\includegraphics[width=0.8\columnwidth]{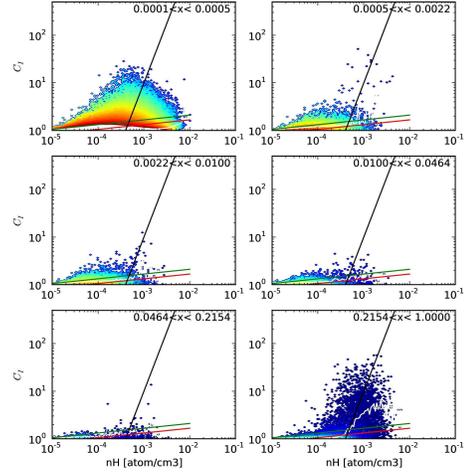}} 
\caption{Clumping factors $C_I$ as a function of the density computed from $8\times8\times8$ cells in the 12.5 Mpc/h simulation and used in the 100 Mpc/h experiment. The six panels show the $C_I$ distributions at different ionization levels~: the solid line shows the $\langle C_I \rangle (n_H)$ trend while the green dotted and the red dashed line stand for $C_I \sim n_H^{0.1}$ models with respectively a high and low normalization. The black dashed line stand for models with $C_I\sim n_H^{2.5}$ at high densities. } 
\label{f:clumpingI} 
\end{figure}

All the $8\times8\times8$ cubes in the 12.5 Mpc/h simulation are considered in the present analysis. Clumping factors are computed by averaging the relevant quantities on these 512 cells. This $8^3$ cells volume in the 12.5 Mpc/h corresponds to the volume of one cell in the 100 Mpc/h. The distributions of $C_R$ and $C_I$ as a function of the density $n_H$ are shown in Figure \ref{f:clumpingR} and \ref{f:clumpingI} for six bins of ionization levels. This distributions were performed by averaging 6 snapshots between $13.1<z<5.9$, yielding clumping factors which do not depend on redshift but only on the physical properties of the cell. This choice aims at simplicity and can suffer from two caveats, one statistical, the other more physical. First the distribution can be skewed by a snapshot which dominates the other. Second, by summing the contributions at all redshift we basically ignore a possible redshift dependance of the clumping and somehow ignore the ionization history of a cell. Still, we checked that the models described below still holds on a redshift by redshift basis.

Clearly, the Figures \ref{f:clumpingR} and \ref{f:clumpingI} present an important spread of values around the mean trend (shown in white). It should not come as a surprise since we somehow projected the clumping factors on the $x,n_H$ space, putting aside e.g. the temperature or the local ionization field. Still some trends can be fitted to a good level of approximation. Considering the distributions of $C_R$ first the distribution are reasonably fitted by $C_R\sim n_H^{0.7}$ trends, especially considering ionized fractions greater than 0.005. The fit is poorer for smaller $x$ but such level of neutral gas do not contribute much to recombination. The same normalization can be applied for $0.005<x<0.2$ (see the green dotted curves) but a smaller normalization (red dashed line) seems to be necessary to fit the mean trend for the last class of ionization . Such results are consistent with the `B' clumping factors found by \citet{2007ApJ...657...15K}. Interestingly, we also recover their `A' clumping factors which are biased toward high densities and which follow a $C_R\sim n_H^{2.5}$ power law (shown as dashed dark lines)~: a subsample of cells follow this trend for high densities like the outliers in the $0.04<x<0.2$ panel or the high density rise of the distribution in the $x>0.2$ panel. By looking at distribution for single redshifts (not shown here), $C_R$ coefficients with the same `A' trend can be found for all the ionized fractions $x>0.0005$ at the highest densities. But their overall weight is such that volume averaged trends tend to follow a `B' law with a gentler slope (shown in white), 0.7 instead of 2.5.

From there it is clear that a single normalization of the $C_R\sim n_H^{0.7}$ law cannot be representative of all the ionization fraction classes. Therefore we choose to perform two sets of runs, one with a `high normalization' (shown in green in Figure \ref{f:clumpingR}) and one with a `low normalization' (shown in red). The former is adequate for $x<0.2$ but overestimates clumping in ionized regions while the latter underestimates clumping in regions with high neutral fraction but is a better fit of the clumping in ionized cells.  Again, this is consistent with the result found by \citet{2007ApJ...657...15K} who studied the neutral fraction dependence of the clumping factors.  Furthermore and as shown hereafter, no strong differences can be noted between these two calculations suggesting that a more detailed $C_R$ with an $x$ dependance (which should lie in between) would lead to similar results.

Considering next the photoionization clumping $C_I$, which distributions are shown in Figure \ref{f:clumpingI}, it clearly appear that the clumping is less pronounced than for recombination. The mean trend can be fitted by $C_R\sim n_H^{0.2}$ laws (shown as red and green lines in the panels), which we used in the subsequent experiments. It should be noted that we recover again a $C_I \sim n_H^{2.5}$ for high density outliers but they are not representative of the overall distributions of $C_I$ values	. In the end, it appears that the photoionization clumping factors are much smaller than recombination one and incidentally during this work we performed calculations with $C_I=1$ (and $C_R\sim n_H^{2.5}$). It is equivalent to assuming an homogeneous UV background~: from the Figure \ref{f:J21nh} it is quite clear that $J_{21}$ does not depend strongly on the density for a large range of values and the approximation made by choosing $C_I=1$ should be reasonably close to the actual clustering.

\subsubsection{Results} 

\begin{figure}
\center{
\includegraphics[width=\columnwidth]{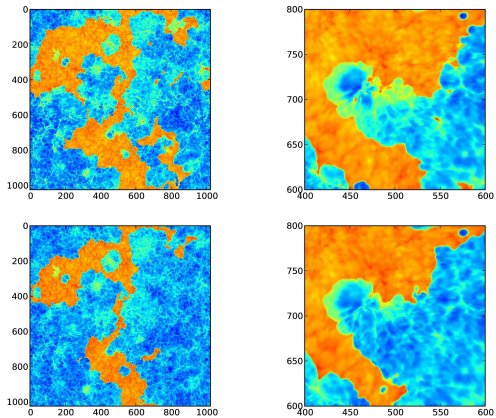}} 
\caption{Neutral fraction maps (red zones are neutral, blue ones ionized) in a
  slice of thickness equals to 9.7 kpc/h comoving at z=6.97. Distances are given in pixels and each map side covers 100 Mpc/h comoving (left column) and 19.53 Mpc/h comoving (right column). The top row pictures stand for the calculations with subgrid clumping, the bottom one for the same calculation without it.} 
\label{f:mapclump} 
\end{figure}

The simple clumping models described in the previous section were added to the basic version of the chemistry/cooling module, and radiative transfer has been performed again for the 100 Mpc/h box, with the same boost factor as the one given in Table \ref{t:boost}. The simple visual inspection of the fields is quite informative on the impact of our subgrid clumping model on transfer calculations~: Figure \ref{f:mapclump} presents the same slice within the box, at the same instant during the pre-overlap phase but for calculations with or without subgrid model. First the overall geometry can be recognized in both calculations, however the experiment with subgrid clumping model presents less extended ionized regions indicating that the overall chronometry has been modified~: with clumping the radiation field is less efficient in ionizing the gas and requires therefore a longer time to achieve a certain level of ionization. Moreover, the experiment without clumping present ionization front which appear smoother than they are in the subgrid clumping model, reflecting again the increased difficulty for radiation to pass through higher density regions. Finally, if one looks closer at photo-ionized regions, much more pockets of neutral gas are seen in the clumping model, as a consequence of the larger recombination rate.

\begin{figure*}
[htbp] \centering 
\includegraphics[width=1.5\columnwidth]{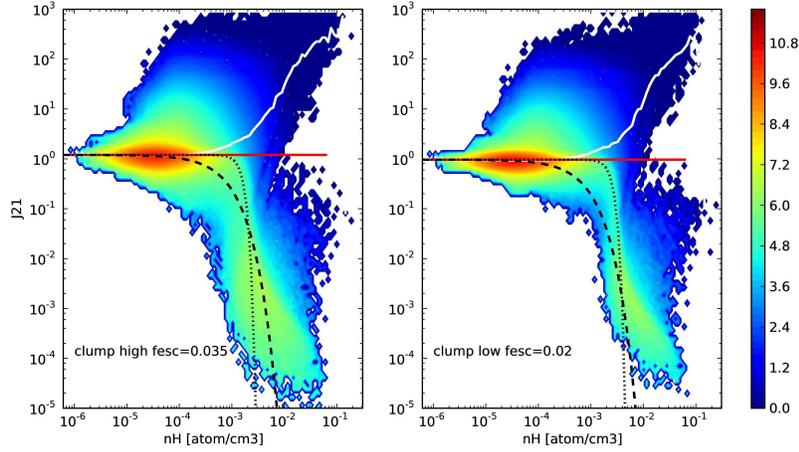} 
\caption{Density dependance of the ionizing intensity in the 100 Mpc/h box
  with subgrid clumping. The red dashed line shows our constant ionizing background model $J_0$, the dashed black line our $J_{21}=J_0\exp(-n_H/n_1^*)$ model and the dotted black line our $J_{21}=J_0\exp(-(n_H/n_4^*)^4)$ model. The white lines show the average intensity per density bin. The top row results were obtained assuming the clumping law with high normalization and the bottom one with low normalization. } 
\label{fig:plot_dense_clump} 
\end{figure*}

\begin{figure*}
[htbp] \centering 
\includegraphics[width=1.5\columnwidth]{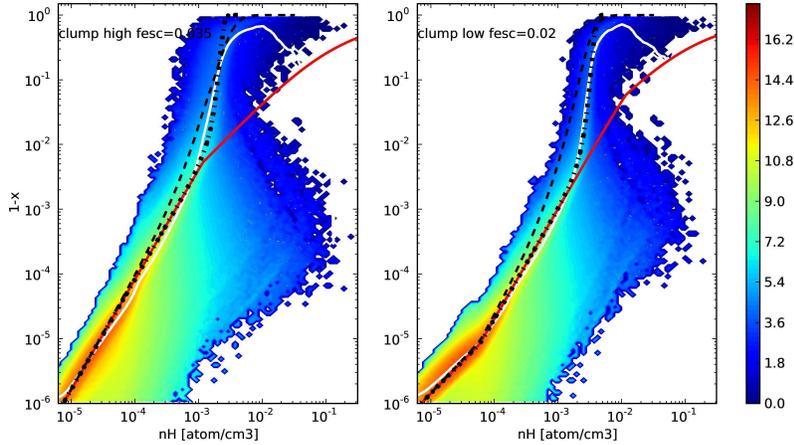} 
\caption{Density dependance of the neutral fraction in the 100 Mpc/h box with
  subgrid clumping. The lines show the neutral fraction for our three ionizing background models, assuming photoionization equilibrium and a clumping factor model. The white lines show the average neutral fraction per density bin. The top row results were obtained assuming the clumping law with high normalization and the bottom one with low normalization. } 
\label{fig:plot_densx_clump} 
\end{figure*}

To assess more quantitatively these aspects, we present in Figure \ref{fig:plot_dense_clump} the same distributions as in section \ref{s:densdep} namely $x(n_H)$ and $J_{21}(n_H)$ but within our clumping factor model. These distributions are given at a post-overlap redshift ($z=5.92$). Like previously, the volume-averaged radiation field follows closely the flux in low density regions  ($n_H<5\times 10^{-4}$ cm$^{-3}$), where its intensity is quasi-constant. From this density upward, a strong exponential cutoff is observed, with a radiation field three orders of magnitude smaller than the average value. Clearly, high density regions live in a radiation field different than the rest of the simulated volume. We use the same type of models $J_0$, $J_1$ and $J_4$ as previously with $n_1^*=0.006$ cm$^{-3}$ for both clumping models and $n_4^*=0.016-0.025$ cm$^{-3}$ for resp. the high and low normalisation models. We recompute the equilibrium ionized fraction and compare it to the distribution actually found in the numerical experiment. The calculations are performed assuming the same clumping models as the one used during the simulation and shown in Figure \ref{fig:plot_densx_clump}. When compared to the calculation without subgrid clumping, it can be noted that the fraction of neutral is more important and that gas tend to be more neutral at a given density. The "cobra rise" of the average $J_{21}$ as a function of $z$ is steeper with clumping and saturates at lower density than it used to. At density close to $5\times 10^{-2}$ cm$^{-3}$, the distribution of neutral fraction is clearly bimodal: a first peak stands for high density regions which are ionized $1-x\sim 0.001$ while a second population has $1-x\sim 1$. It indicates that some gaseous regions are sufficiently embedded and/or recombine fast enough to be "spared" by the radiation field. Again the $J_0$ model is completely off the mean $1-x(n_H)$ trend for high density regions $n_H>0.001$ cm$^{-3}$, even though this level of radiation is effectively the one measured when averaging over the whole volume. Meanwhile the $J_1$ does a better job and $J_4$ appears to be a good match which is not surprising since they are better fits to the local density dependance of the UV background. Because of self-shielding, it is clear that the ionized state of high density regions cannot be deduced by the simple generalization of the average UV field found in the simulated volume.

\subsubsection{Comparison to observational constraints}

\begin{figure}
[htbp] \centering 
\includegraphics[width=1.\columnwidth]{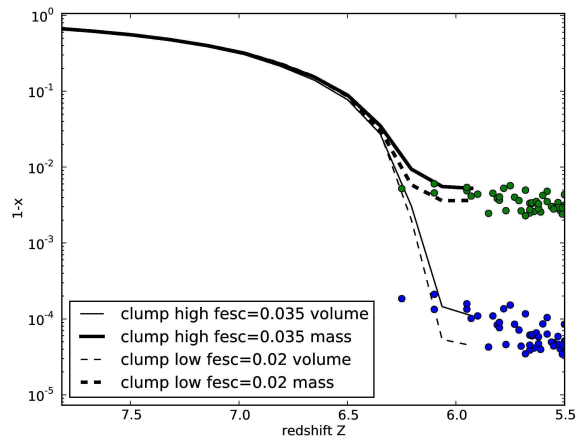} 
\caption{Evolution of the mass and volume averaged neutral fraction in the 100 Mpc/h box with a clumping factor assuming a high/low normalization (thin/thick lines). The values at $z=6$ are consistent with measurements made by \citet{2006AJ....132..117F} for both kind of average method (dots).} 
\label{fig:plot_x_clump_ok} 
\end{figure}

We first consider the evolution of the volume and mass averaged neutral fraction, shown in Figure \ref{fig:plot_x_clump_ok}. Compared to the experiments without subgrid clumping, the fractions are typically one order of magnitude larger when subgrid structures are modeled. At $z=5.9$, the volume averaged neutral fractions are equal to $x\sim5\times 10^{-5}$ and the mass averaged to $x\sim 3\times10^{-3}$. The differences between the low and high normalization models for the clumping are small with higher neutral fraction for the high normalization model, which is expected since it overestimates the recombination rate in the most neutral regions. These levels of neutral gas are consistent with observational constraints provided by \citet{2006AJ....132..117F}\footnote{The observational constraints shown here were recomputed from the transmission tables of \citet{2006AJ....132..117F} using the same cosmology as the one used in our calculations. It results in a relative variation of $20\%$.} and for the two types of averaging and indicates that high resolution or subgrid clumping are required in order to match the data. For instance, the same agreement has already been obtained by \citet{2007ApJ...671....1T} using directly the particles as a source of clumping in a high resolution pure DM simulation. Other examples are \citet{2007ApJ...657...15K} using clumping factors and \citet{2006ApJ...648....1G} using small boxes calculations for the volume weighted neutral fraction. 

The agreement found in the current work is encouraging but should nevertheless considered with some care. First, the clumping models are simple and lack a detailed dependance on e.g. the temperature or UV field. But even with a more accurate description of the subcell physics, clumping factors will remain as a trick to cope with inadequate resolution and the forthcoming effort should concentrate on improving the resolution using larger simulations. Furthermore, we still lack the coupling with the hydrodynamics which is likely to affect to small scale features of reionization and such physics cannot be assessed using only a subgrid clumping model. In the end, we estimate that given the simplistic aspect of our clumping model, the current agreement should be seen as a sign that the overall direction is correct, but it is not clear that improving the model is worth the effort, comparing to increasing the resolution of the simulation. Second, \citet{2007ApJ...671....1T} already noted that the constraints provided by \citet{2006AJ....132..117F} also depend on some rather strong assumptions like the strict uniformity of the UV background or a smooth gas distribution. In the most pessimistic case, the present agreement can be fortuitous, even though an agreement on both the volume and mass averaged neutral fraction is unlikely to happen by accident. 

\begin{figure*}
[htbp] \centering 
\includegraphics[width=1.5\columnwidth]{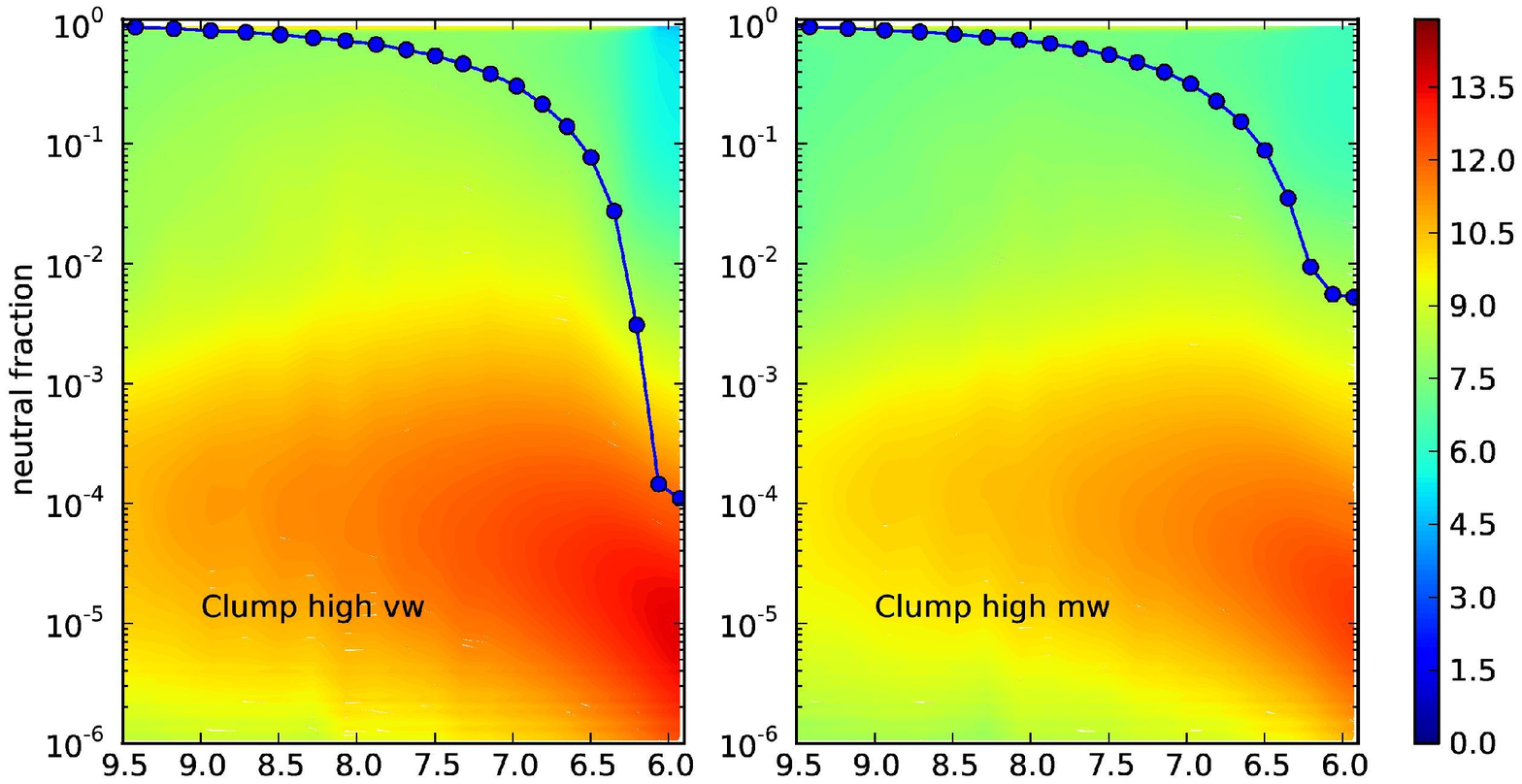} 
\includegraphics[width=1.5\columnwidth]{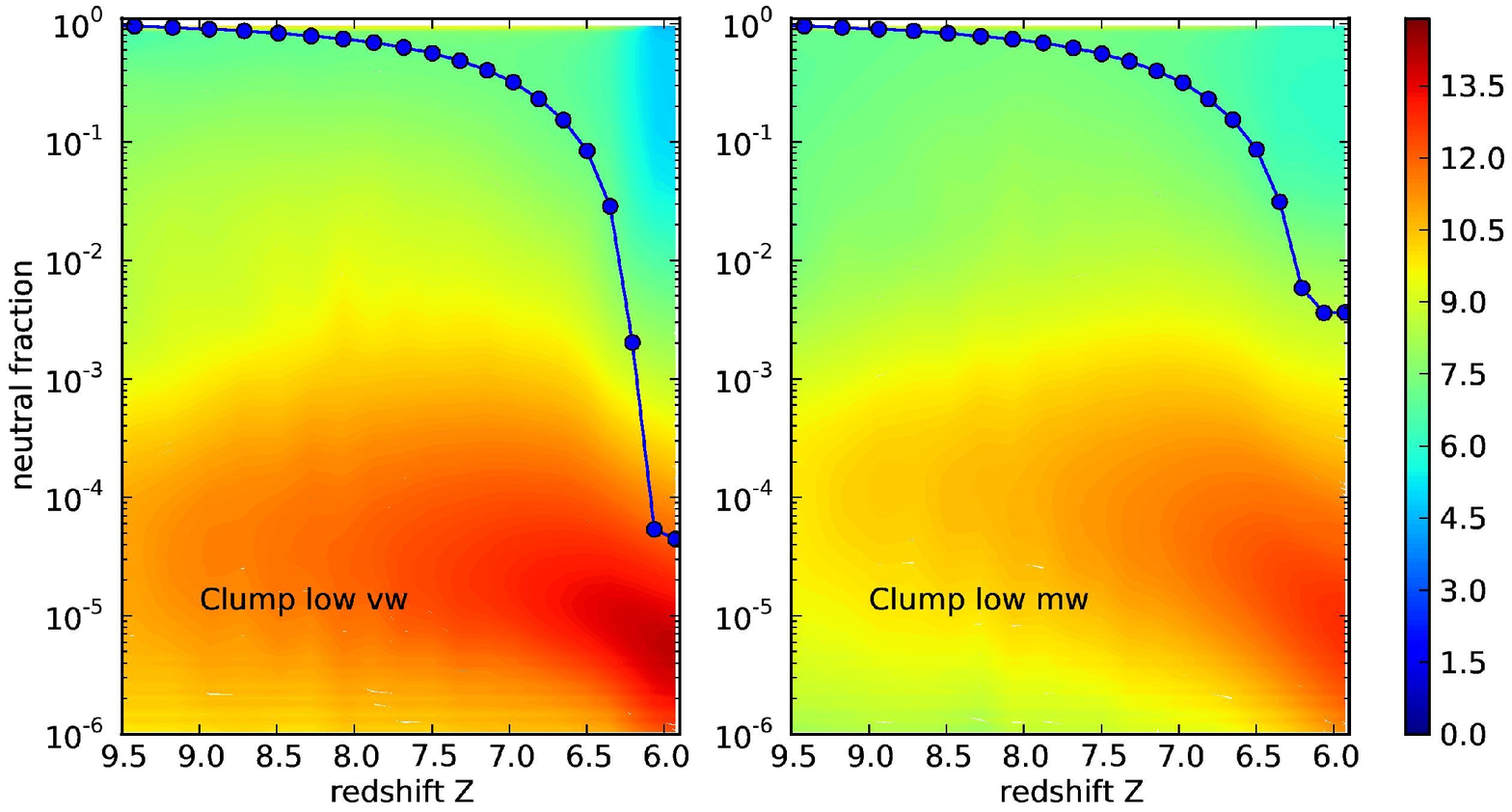} 
\caption{Evolution of the neutral fraction distribution with redshift along with the evolution of the average value using mass weighting (bottom row) and volume weighting (top row) for the 100 Mpc/h box with clumping. The left column stand for experiments with high normalization clumping and the right one for the low normalization clumping model. Contours show the density probabilities of neutral fraction with high probability densities in red and low ones in blue.} \label{fig:plot_histo_x_clump_ok_high} 
\end{figure*}

It should however be noted that the average neutral fractions do not correspond to any feature in the probability density distributions. In Figure \ref{fig:plot_histo_x_clump_ok_high}, we overlay the evolution of the averaged neutral fractions on the evolving distributions of probabilities. Red regions stand for high probabilities while blue ones stand for lower probabilities~: after the overlap, the distributions are clearly dominated by strongly ionized regions ($x\sim 10^{-5}$) which correspond mostly to low gas densities. The mass-weighted distribution enhance the contribution of dense cells and which in turn push the average neutral fraction toward higher values ($x\sim10^{-2}$). However, the values of the average neutral fractions are never coincident with the maximum of the distribution. On the contrary these averages lie in the transition region between low neutral fraction and high neutral fraction regions. In other words, the average values lie in-between two characteristic regions of the gas distribution, but is representative of neither of them and consequently is not a good proxy of the physical states that coexist inside the simulation.

\begin{figure}
[htbp] \centering 
\includegraphics[width=0.8\columnwidth]{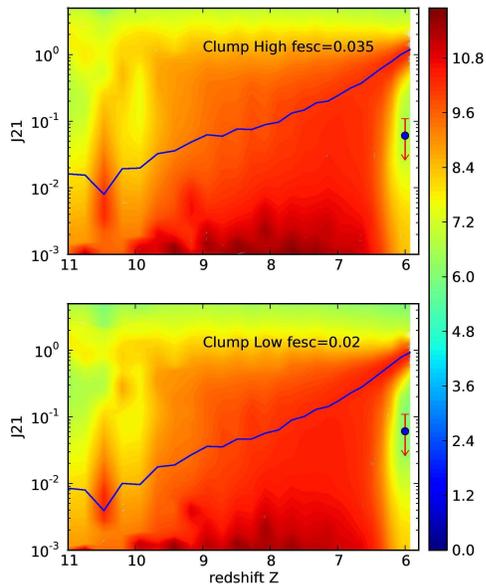} 
\caption{Evolution of the average ionizing radiation (blue line) in the 100 Mpc/h box assuming a subgrid clumping factor model with a low/high normalization(top/bottom panel). The colored isocontours stand for the distribution of $J_{21}$ values at each redshift with high probability densities in red and low ones in blue. The marker at $z\sim 6$ shows the constraint provided by \citet{2007MNRAS.382..325B}. } 
\label{fig:plot_J21_clump} 
\end{figure}

In Figure \ref{fig:plot_J21_clump}, we present the evolution of the distribution of the UV background in the 100 Mpc/h boxes with subgrid clumping with the mean value superimposed and the constraint provided by \citet{2007MNRAS.382..325B} after the overlap. Compared to the same calculation without clumping, some improvement can be seen and the intensity of the UV background has been reduced by a factor 2 or 3 (depending on the kind of normalization applied to the recombination clumping). It can be noted that the same ratio was already observed between the 100 Mpc/h and the 12.5 Mpc/h (which served to calibrate our clumping model) boxes in our fiducial calculations. Still, the discrepancy with the observational constraints remains quite large, almost one order of magnitude. Furthermore, the inspection of the distribution indicates that the mean value of the UV background effectively tracks the maximum of the $J_{21}$ distribution. Therefore, no bias or multimodal distribution can be invoked as a valid reason for the discrepancy.

This failure can be explained by the fact that very different regions are at the origin of the average value for the neutral fraction and for the UV background values. After the overlap, the neutral fraction is intrinsically low in low density regions ($1-x\sim10^{-5}-10^{-6}$) and few regions with high neutral fraction push the average to higher values~: at face value a single fully neutral cell weighs as much as $10^6$ cells with a $10^{-6}$ neutral fraction. Furthermore, if mass-weighting is considered, the impact of such cells is even higher since they are usually more neutral. As a consequence, the average neutral fraction is pushed toward values higher than the peak of the distribution and dense cells (even less numerous) have an important impact. Considering now the average UV background, dense cells lie in regions where the radiation intensity is typically one thousand times smaller than the typical value computed in low density regions because of self-shielding, which implies a smaller impact on the average $J_{21}$. Consequently, the mean neutral fraction is not related to the mean UV background~: the former is influenced by dense clumps while the latter is mainly set by voids. The fact that dense regions do not lie in the typical UV background explains our ability to reproduce the neutral fraction and our failure to satisfy the constraints on the ionizing radiation field.

One may therefore ask how do we balance a discrepancy in the photoionization rate and an agreement in neutral fraction? First let us recall that the actual quantity measured in quasar spectra is the transmission $\mathcal{T}$, i.e. the ratio of the observed flux to the unobscured one over a given range of redshifts \citep[see e.g.][]{2002AJ....123.1247F, 2006AJ....132..117F}. At the redshift considered here, the equivalent comoving distances are of the order of $60$ Mpc/h and are therefore comparable to the experiments presented in the current work. Hence, observations gives a constraint on : 
\begin{equation}
\langle \mathcal{T} \rangle \sim \int p(\Delta) e^{-\alpha Q(z)\Delta^2/\Gamma} d\Delta, \label{eq:trans} 
\end{equation}
where $\Delta$ stands for the density contrast, $\alpha$ is the recombination rate (mostly homogeneous), Q(z) depends on the physical parameters of the Ly-$\alpha$ transmission and cosmology and finally $p(\Delta)$ is the PDF of the density. A typical example of such a PDF is given by \citet{2000ApJ...530....1M} which is used in the models of \citet{2002AJ....123.1247F}, \citet{2006AJ....132..117F} or \citet{2007MNRAS.382..325B}. In Eq. \ref{eq:trans}, the photoionization rate can be deduced from $\langle T \rangle$ and assuming photoionization equilibrium the neutral fraction can also be deduced. We have seen that the latter assumption is mostly verified. The relation also assumes that the Universe is mostly ionized and that the UV background is homogeneous. From the expression in Equation~\ref{eq:trans}, it can easily be seen that the exponential cutoff implies that the actual density distribution at high density has no influence on the observed quantity, therefore their departure from homogeneity and low neutral fraction (measured in simulations) does not impact on the transmissions. Conversely, it implies that the quantities which are mostly constrained are inferred from low-density regions \citep[for a detailed discussion see e.g.][]{2005ApJ...620L...9O}.  In other words the photoionisation rate is more `reliable' or more directly constrained than the neutral fraction and should be reproduced first by simulations~: in principle the agreement on the neutral fraction should follow. In the current work, we show however that reproducing first the neutral fraction does not automatically imply an adequate photoionisation rate. 
 
\subsubsection{Improving the model} 

\begin{figure*}
[htbp] \centering 
\includegraphics[width=1.5\columnwidth]{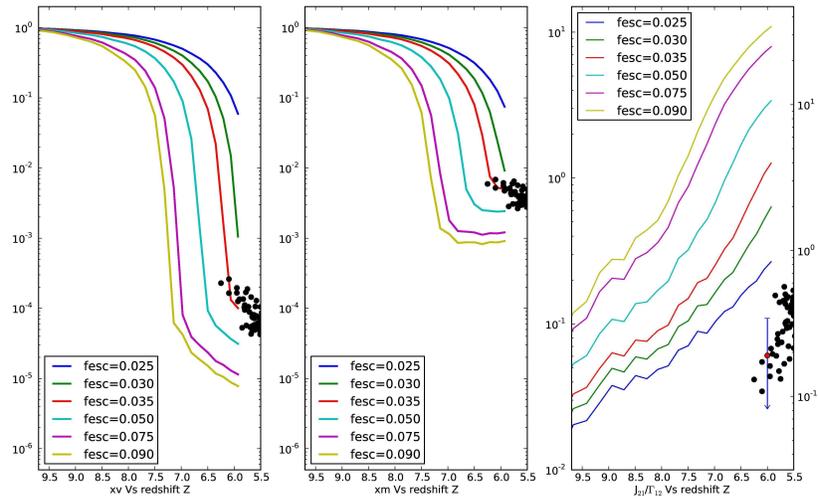} 
\caption{Evolution of the mass, volume averaged neutral fraction and ionizing rate in the 100 Mpc/h box with a clumping factor assuming a high normalization for various escape fractions. The values at $z=6$ are consistent with measurements made by \citet{2006AJ....132..117F} (dots). The red arrow at $z \sim 6$ shows the constraint provided by \citet{2007MNRAS.382..325B}.} 
\label{fig:plot_x_clump_ok_fesc} 
\end{figure*}

Which path should be taken toward a complete agreement between observational constraints and our calculations? The most obvious free parameter we have access to is the escape fraction. We present in Figure~\ref{fig:plot_x_clump_ok_fesc} the evolution of the averaged UV background and neutral fraction for various escape fractions and using the same $1024^3$ particles 100 Mpc/h simulation with the high normalization clumping factor model. Plotted along are the constraints provided by \citet{2006AJ....132..117F} and \citet{2007MNRAS.382..325B}. As expected, lowering the escape fraction make the simulated UV background more consistent with observations. However, also as expected, the redshift of reionization decreases and for the lowest value of $f_\mathrm{esc}=2.5\%$ presented here, overlap is not complete and the average neutral fraction is only at $5\%$ at $z=6$. Such a scenario is problematic, because it implies that the neutral fraction must decrease very sharply~: observational data exhibit some transmission for quasars at redshifts $z\sim 5.9$, i.e. at levels of neutral fraction close to $1-x\sim 0.0001$ and it would imply a sudden decrease from $x\sim0.1$ to $10^{-4}$ in a small redshift interval of $\Delta z\sim0.1$. Furthermore such a trend would also go against a better agreement with the optical depth measured from the CMB data which favor a higher escape fraction. One option would be to use an evolving escape fraction, from $\sim 10\%$ at $z\sim 10$ fown to $f_\mathrm{esc}\sim 1\%$ at $z\sim 6$. Preliminary experiments (not shown here) indicate that, albeit helpful, this option does not easily provide a solution to the discrepancy. Furthermore, even if a good match is obtained, it would only consist in a proof of concept and one would have to relate this evolution to a physical process (like e.g. star formation). Other routes can be used to reduce this discrepancy. The lack of multi-frequency transfer implies among other things that no preheating occur behind ionization fronts. In particular, it would reduce both the recombination rate of the gas and the required number of photons per baryons to complete reionization. Finally, the proper coupling of radiative transfer with hydrodynamics may prove to be crucial~: low density regions or minihaloes are likely to react to any kind of heating due to radiative transfer and the source production (namely star formation) may be affected \citep{2005MNRAS.361..405I, 2007MNRAS.376..534I}, even though self-shielding, which has been shown to be quite effective in our calculations, could go in the opposite direction. These points will be investigated in future work.

\section{Summary and Prospects} 

We have presented a set of radiative cosmological simulations in order to model the reionization epoch from  $z \sim 18$ down to $z \sim 6$. The gas and dark matter dynamics, as well as the associated star formation have been performed with the {\tt RAMSES} code, while radiative transfer has been computed by means of a moment--based formalism using the M1 closure relation, implemented in the {\tt ATON} code. The latter has been ported on a multi-GPU architecture using CUDA, providing an acceleration close to 100x, which allows us to tackle radiative transfer problems at high resolution (a $1024^3$ base grid and 2 levels of refinement for the hydrodynamics and  a $1024^3$ Cartesian grid for the radiative transfer). 

A good level of convergence on average quantities (neutral fraction, UV background and Thomson optical depth) has been observed between different simulations of increasing mass and spatial resolution, as long as the effect of finite mass resolution on the simulated star formation history are properly taken in account. We have also shown that the density dependance of the neutral fraction is close to the one predicted by photo-ionization equilibrium, as long as the effect of self-shielding are considered when defining the properties of the UV field. It also appears that without any other ingredients, our simulation fails at reproducing the $z\sim 6$ constraints on the neutral fraction of hydrogen and the intensity of the UV background, in a similar manner to \citet{2009MNRAS.400.1049F}.

By combining our best resolved simulation (12.5 Mpc/h and 1024$^3$ particles) with our largest simulated volume (100 Mpc/h with 1024$^3$ particles), we have introduced a subgrid clumping model in our chemistry solver, consistent with the one derived by e.g. \citet{2007ApJ...657...15K}. We have shown that, although this clumping factor model is quite simplistic, its allowed us to reproduce the level of neutral gas deduced from the spectra of high redshift quasars, as did previously \citet{2006ApJ...648....1G} or \citet{2007ApJ...671....1T} among others. However, our estimation of the average photoionization rate is still at least a factor of 2 above the observational constraints. This "semi-success" can be explained by the fact that the average radiation intensity and the average neutral fraction depend on different regions of the gas distribution and one cannot simply deduce one from the other using photoionization equilibrium~: in other words, if one constraint is satisfied, the other can't be. However, we have argued that the photoionization rate is probably a more robust observational constraint than the neutral fraction. This suggests that some effort should still be done in our modelisation to reproduce the level of the UV background at $z\sim 6$.

Among several prospects, one obviously think of increasing the resolution of the calculations. With GPU acceleration $2048^3$ hydro+ radiative transfer calculations are within reach. However, it clearly appears that coupled hydrodynamics and radiative transfer simulations are necessary at this stage, since an increase in resolution will inevitably rise the question of the impact of radiation on mini-haloes or on the star formation history. Also additional physics should be implemented, such as multi-group radiative transfer, where the importance of preheating by X-rays could therefore be fully assessed (see e.g. \citet{2006MNRAS.371..867F, 2008ApJ...685....1S}), but also population III stars (\citet{2007ApJ...671....1T}) and  varying star formation efficiencies and escape fractions (\citet{2008ApJ...672..765G,2009ApJ...693..984W}). Overall, on a final positive note, our current results indicate a satisfying trend of cosmological calculations toward satisfying observational constraints.

\noindent
{\bf Acknowledgments.}
DA is supported by the ANR grant LIDAU and a \emph{Conseil Scientifique} Grant from the University of Strasbourg.
This work was granted access to the HPC resources of CCRT under the "Grand Challenge Applications" allocation for 2009. 

\bibliography{aubert_teyssier_2010} 

\begin{thebibliography}{57}
\expandafter\ifx\csname natexlab\endcsname\relax\def\natexlab#1{#1}\fi

\bibitem[{{Abel} {et~al.}(1999){Abel}, {Norman}, \&
  {Madau}}]{1999ApJ...523...66A}
{Abel}, T., {Norman}, M.~L., \& {Madau}, P. 1999, \apj, 523, 66

\bibitem[{{Anninos} {et~al.}(1997){Anninos}, {Zhang}, {Abel}, \&
  {Norman}}]{1997NewA....2..209A}
{Anninos}, P., {Zhang}, Y., {Abel}, T., \& {Norman}, M.~L. 1997, New Astronomy,
  2, 209

\bibitem[{Aubert {et~al.}(2009)Aubert, Amini, \& David}]{1560857}
Aubert, D., Amini, M., \& David, R. 2009, in ICCS '09: Proceedings of the 9th
  International Conference on Computational Science (Berlin, Heidelberg:
  Springer-Verlag), 874--883

\bibitem[{{Aubert} \& {Teyssier}(2008)}]{2008MNRAS.387..295A}
{Aubert}, D., \& {Teyssier}, R. 2008, \mnras, 387, 295

\bibitem[{{Baek} {et~al.}(2009){Baek}, {Di Matteo}, {Semelin}, {Combes}, \&
  {Revaz}}]{2009A&A...495..389B}
{Baek}, S., {Di Matteo}, P., {Semelin}, B., {Combes}, F., \& {Revaz}, Y. 2009,
  \aap, 495, 389

\bibitem[{{Barkana} \& {Loeb}(2001)}]{2001PhR...349..125B}
{Barkana}, R., \& {Loeb}, A. 2001, \physrep, 349, 125

\bibitem[{{Bertschinger}(1998)}]{1998ARA&A..36..599B}
{Bertschinger}, E. 1998, \araa, 36, 599

\bibitem[{{Bolton} \& {Haehnelt}(2007)}]{2007MNRAS.382..325B}
{Bolton}, J.~S., \& {Haehnelt}, M.~G. 2007, \mnras, 382, 325

\bibitem[{{Cen}(1992)}]{1992ApJS...78..341C}
{Cen}, R. 1992, \apjs, 78, 341

\bibitem[{{Ciardi} {et~al.}(2001){Ciardi}, {Ferrara}, {Marri}, \&
  {Raimondo}}]{2001MNRAS.324..381C}
{Ciardi}, B., {Ferrara}, A., {Marri}, S., \& {Raimondo}, G. 2001, \mnras, 324,
  381

\bibitem[{Dubroca \& Feugeas(1999)}]{Dubroca1999915}
Dubroca, B., \& Feugeas, J.-L. 1999, Comptes Rendus de l'Academie des Sciences
  - Series I - Mathematics, 329, 915

\bibitem[{{Efstathiou} {et~al.}(1985){Efstathiou}, {Davis}, {White}, \&
  {Frenk}}]{1985ApJS...57..241E}
{Efstathiou}, G., {Davis}, M., {White}, S.~D.~M., \& {Frenk}, C.~S. 1985,
  \apjs, 57, 241

\bibitem[{{Fan} {et~al.}(2002){Fan}, {Narayanan}, {Strauss}, {White}, {Becker},
  {Pentericci}, \& {Rix}}]{2002AJ....123.1247F}
{Fan}, X., {Narayanan}, V.~K., {Strauss}, M.~A., {White}, R.~L., {Becker},
  R.~H., {Pentericci}, L., \& {Rix}, H. 2002, \aj, 123, 1247

\bibitem[{{Fan} {et~al.}(2006){Fan}, {Strauss}, {Becker}, {White}, {Gunn},
  {Knapp}, {Richards}, {Schneider}, {Brinkmann}, \&
  {Fukugita}}]{2006AJ....132..117F}
{Fan}, X., {et~al.} 2006, \aj, 132, 117

\bibitem[{{Finlator} {et~al.}(2009{\natexlab{a}}){Finlator}, {{\"O}zel}, \&
  {Dav{\'e}}}]{2009MNRAS.393.1090F}
{Finlator}, K., {{\"O}zel}, F., \& {Dav{\'e}}, R. 2009{\natexlab{a}}, \mnras,
  393, 1090

\bibitem[{{Finlator} {et~al.}(2009{\natexlab{b}}){Finlator}, {{\"O}zel},
  {Dav{\'e}}, \& {Oppenheimer}}]{2009MNRAS.400.1049F}
{Finlator}, K., {{\"O}zel}, F., {Dav{\'e}}, R., \& {Oppenheimer}, B.~D.
  2009{\natexlab{b}}, \mnras, 400, 1049

\bibitem[{Fromang {et~al.}(2006)Fromang, Hennebelle, \&
  Teyssier}]{Fromang:2006p400}
Fromang, S., Hennebelle, P., \& Teyssier, R. 2006, Astronomy and Astrophysics,
  457, 371

\bibitem[{{Furlanetto}(2006)}]{2006MNRAS.371..867F}
{Furlanetto}, S.~R. 2006, \mnras, 371, 867

\bibitem[{Gnedin(2000)}]{Gnedin:2000p1555}
Gnedin, N.~Y. 2000, The Astrophysical Journal, 542, 535

\bibitem[{{Gnedin} \& {Abel}(2001)}]{2001NewA....6..437G}
{Gnedin}, N.~Y., \& {Abel}, T. 2001, New Astronomy, 6, 437

\bibitem[{{Gnedin} \& {Fan}(2006)}]{2006ApJ...648....1G}
{Gnedin}, N.~Y., \& {Fan}, X. 2006, \apj, 648, 1

\bibitem[{{Gnedin} \& {Hui}(1998)}]{1998MNRAS.296...44G}
{Gnedin}, N.~Y., \& {Hui}, L. 1998, \mnras, 296, 44

\bibitem[{{Gnedin} {et~al.}(2008){Gnedin}, {Kravtsov}, \&
  {Chen}}]{2008ApJ...672..765G}
{Gnedin}, N.~Y., {Kravtsov}, A.~V., \& {Chen}, H. 2008, \apj, 672, 765

\bibitem[{{Gonz{\'a}lez} {et~al.}(2007){Gonz{\'a}lez}, {Audit}, \&
  {Huynh}}]{2007A&A...464..429G}
{Gonz{\'a}lez}, M., {Audit}, E., \& {Huynh}, P. 2007, \aap, 464, 429

\bibitem[{Governato {et~al.}(2009)Governato, Brook, Brooks, Mayer, Willman,
  Jonsson, Stilp, Pope, Christensen, Wadsley, \& Quinn}]{Governato:2009p1455}
Governato, F., {et~al.} 2009, Monthly Notices of the Royal Astronomical
  Society, 398, 312

\bibitem[{Governato {et~al.}(2010)Governato, Brook, Mayer, Brooks, Rhee,
  Wadsley, Jonsson, Willman, Stinson, Quinn, \& Madau}]{Governato:2010p1442}
---. 2010, Nature, 463, 203, (c) 2010: Nature

\bibitem[{{Hernquist} {et~al.}(1991){Hernquist}, {Bouchet}, \&
  {Suto}}]{1991ApJS...75..231H}
{Hernquist}, L., {Bouchet}, F.~R., \& {Suto}, Y. 1991, \apjs, 75, 231

\bibitem[{Hoeft {et~al.}(2006)Hoeft, Yepes, Gottl{\"o}ber, \&
  Springel}]{Hoeft:2006p1565}
Hoeft, M., Yepes, G., Gottl{\"o}ber, S., \& Springel, V. 2006, Monthly Notices
  of the Royal Astronomical Society, 371, 401

\bibitem[{{Iliev} {et~al.}(2006{\natexlab{a}}){Iliev}, {Mellema}, {Pen},
  {Merz}, {Shapiro}, \& {Alvarez}}]{2006MNRAS.369.1625I}
{Iliev}, I.~T., {Mellema}, G., {Pen}, U., {Merz}, H., {Shapiro}, P.~R., \&
  {Alvarez}, M.~A. 2006{\natexlab{a}}, \mnras, 369, 1625

\bibitem[{{Iliev} {et~al.}(2007){Iliev}, {Mellema}, {Shapiro}, \&
  {Pen}}]{2007MNRAS.376..534I}
{Iliev}, I.~T., {Mellema}, G., {Shapiro}, P.~R., \& {Pen}, U. 2007, \mnras,
  376, 534

\bibitem[{{Iliev} {et~al.}(2005){Iliev}, {Shapiro}, \&
  {Raga}}]{2005MNRAS.361..405I}
{Iliev}, I.~T., {Shapiro}, P.~R., \& {Raga}, A.~C. 2005, \mnras, 361, 405

\bibitem[{{Iliev} {et~al.}(2006{\natexlab{b}}){Iliev}, {Ciardi}, {Alvarez},
  {Maselli}, {Ferrara}, {Gnedin}, {Mellema}, {Nakamoto}, {Norman}, {Razoumov},
  {Rijkhorst}, {Ritzerveld}, {Shapiro}, {Susa}, {Umemura}, \&
  {Whalen}}]{2006MNRAS.371.1057I}
{Iliev}, I.~T., {et~al.} 2006{\natexlab{b}}, \mnras, 371, 1057

\bibitem[{{Iliev} {et~al.}(2009){Iliev}, {Whalen}, {Mellema}, {Ahn}, {Baek},
  {Gnedin}, {Kravtsov}, {Norman}, {Raicevic}, {Reynolds}, {Sato}, {Shapiro},
  {Semelin}, {Smidt}, {Susa}, {Theuns}, \& {Umemura}}]{2009MNRAS.400.1283I}
---. 2009, \mnras, 400, 1283

\bibitem[{{Katz} {et~al.}(1996){Katz}, {Weinberg}, \&
  {Hernquist}}]{1996ApJS..105...19K}
{Katz}, N., {Weinberg}, D.~H., \& {Hernquist}, L. 1996, \apjs, 105, 19

\bibitem[{{Kohler} {et~al.}(2007){Kohler}, {Gnedin}, \&
  {Hamilton}}]{2007ApJ...657...15K}
{Kohler}, K., {Gnedin}, N.~Y., \& {Hamilton}, A.~J.~S. 2007, \apj, 657, 15

\bibitem[{{Komatsu} {et~al.}(2009){Komatsu}, {Dunkley}, {Nolta}, {Bennett},
  {Gold}, {Hinshaw}, {Jarosik}, {Larson}, {Limon}, {Page}, {Spergel},
  {Halpern}, {Hill}, {Kogut}, {Meyer}, {Tucker}, {Weiland}, {Wollack}, \&
  {Wright}}]{2009ApJS..180..330K}
{Komatsu}, E., {et~al.} 2009, \apjs, 180, 330

\bibitem[{Mayer {et~al.}(2008)Mayer, Governato, \& Kaufmann}]{Mayer:2008p1478}
Mayer, L., Governato, F., \& Kaufmann, T. 2008, Advanced Science Letters, 1, 7

\bibitem[{{McQuinn} {et~al.}(2007){McQuinn}, {Lidz}, {Zahn}, {Dutta},
  {Hernquist}, \& {Zaldarriaga}}]{2007MNRAS.377.1043M}
{McQuinn}, M., {Lidz}, A., {Zahn}, O., {Dutta}, S., {Hernquist}, L., \&
  {Zaldarriaga}, M. 2007, \mnras, 377, 1043

\bibitem[{{Miralda-Escud{\'e}} {et~al.}(2000){Miralda-Escud{\'e}}, {Haehnelt},
  \& {Rees}}]{2000ApJ...530....1M}
{Miralda-Escud{\'e}}, J., {Haehnelt}, M., \& {Rees}, M.~J. 2000, \apj, 530, 1

\bibitem[{{Oh} \& {Furlanetto}(2005)}]{2005ApJ...620L...9O}
{Oh}, S.~P., \& {Furlanetto}, S.~R. 2005, \apjl, 620, L9

\bibitem[{{Prunet} {et~al.}(2008){Prunet}, {Pichon}, {Aubert}, {Pogosyan},
  {Teyssier}, \& {Gottloeber}}]{2008ApJS..178..179P}
{Prunet}, S., {Pichon}, C., {Aubert}, D., {Pogosyan}, D., {Teyssier}, R., \&
  {Gottloeber}, S. 2008, \apjs, 178, 179

\bibitem[{{Rasera} \& {Teyssier}(2006)}]{2006A&A...445....1R}
{Rasera}, Y., \& {Teyssier}, R. 2006, \aap, 445, 1

\bibitem[{{Razoumov} {et~al.}(2002){Razoumov}, {Norman}, {Abel}, \&
  {Scott}}]{2002ApJ...572..695R}
{Razoumov}, A.~O., {Norman}, M.~L., {Abel}, T., \& {Scott}, D. 2002, \apj, 572,
  695

\bibitem[{Schaye \& Vecchia(2008)}]{Schaye:2008p1393}
Schaye, J., \& Vecchia, C.~D. 2008, Monthly Notices of the Royal Astronomical
  Society, 383, 1210

\bibitem[{Sengupta {et~al.}(2007)Sengupta, Harris, Zhang, \&
  Owens}]{Sengupta:2007:SPF}
Sengupta, S., Harris, M., Zhang, Y., \& Owens, J.~D. 2007, in Graphics Hardware
  2007, 97--106

\bibitem[{{Shin} {et~al.}(2008){Shin}, {Trac}, \& {Cen}}]{2008ApJ...681..756S}
{Shin}, M., {Trac}, H., \& {Cen}, R. 2008, \apj, 681, 756

\bibitem[{{Shull} \& {Venkatesan}(2008)}]{2008ApJ...685....1S}
{Shull}, J.~M., \& {Venkatesan}, A. 2008, \apj, 685, 1

\bibitem[{{Songaila}(2004)}]{2004AJ....127.2598S}
{Songaila}, A. 2004, \aj, 127, 2598

\bibitem[{Springel \& Hernquist(2003)}]{Springel:2003p1288}
Springel, V., \& Hernquist, L. 2003, Astrophysical Supercomputing using
  Particle Simulations, 208, 273

\bibitem[{Stinson {et~al.}(2006)Stinson, Seth, Katz, Wadsley, Governato, \&
  Quinn}]{Stinson:2006p1402}
Stinson, G., Seth, A., Katz, N., Wadsley, J., Governato, F., \& Quinn, T. 2006,
  Monthly Notices of the Royal Astronomical Society, 373, 1074

\bibitem[{{Teyssier}(2002)}]{2002A&A...385..337T}
{Teyssier}, R. 2002, \aap, 385, 337

\bibitem[{Teyssier {et~al.}(2006)Teyssier, Fromang, \&
  Dormy}]{Teyssier:2006p413}
Teyssier, R., Fromang, S., \& Dormy, E. 2006, Journal of Computational Physics,
  218, 44, elsevier Inc.

\bibitem[{Toro {et~al.}(1994)Toro, Spruce, \& Speares}]{Toro:1994p1151}
Toro, E.~F., Spruce, M., \& Speares, W. 1994, Shock Waves, 4, 25, (c) 1994:
  Springer-Verlag

\bibitem[{{Trac} \& {Cen}(2007)}]{2007ApJ...671....1T}
{Trac}, H., \& {Cen}, R. 2007, \apj, 671, 1

\bibitem[{{Trac} \& {Gnedin}(2009)}]{2009arXiv0906.4348T}
{Trac}, H., \& {Gnedin}, N.~Y. 2009, ArXiv e-prints

\bibitem[{{Wise} \& {Cen}(2009)}]{2009ApJ...693..984W}
{Wise}, J.~H., \& {Cen}, R. 2009, \apj, 693, 984

\bibitem[{Yepes {et~al.}(1997)Yepes, Kates, Khokhlov, \&
  Klypin}]{Yepes:1997p1245}
Yepes, G., Kates, R., Khokhlov, A., \& Klypin, A. 1997, Monthly Notices of the
  Royal Astronomical Society, 284, 235

\end{thebibliography}
\bibliographystyle{apj}

\appendix 
\section{On the GPU implementation of {\tt ATON}} This section comment in further details the implementation of {\tt ATON} on Graphics Processing Units (GPU). The whole development has been performed using the version 2.2 of the CUDA extension to the C language, developed by Nvidia for its graphics devices. However this section should be seen as a commentary of the implementation process rather than a full description of the programming details~: the field of GPU programming is currently in full expansion, several standards/programming languages are competing with each other and many specific programming details are likely to be quickly outdated. For these reasons we choose to stick fairly general techniques, comment the suitability of the calculations to multithreaded calculations and provide the general tricks of our development to optimize the performances.

Le us first recall that GPU computing relies on two separate hierarchies: a hierarchy of memories and a hierarchy of tasks. Regarding the first aspect and because GPUs are devices physically separated from the host, they possess their own memory known as the video memory or `global memory' in the CUDA nomenclature. The transfer rate between the host and the GPU is therefore strongly limited by a bus and the best performances can be achieved if all the calculations are performed on the device, i.e. without transfer between the host and the GPU. An ideal situation would be to transfer the initial conditions (ICs) on the device and let it process the calculation on its own with the host acting merely as a driver of the calculation. This constraint is satisfied by the GPU implementation of {\tt ATON}~: the host sends signals to the GPU in a regular fashion to advance the simulation within a time step and from one time step to another but it never actually computes anything on the data. More precisely, the ICs are sent on the GPU, then the host asks to the GPU to compute the radiation transport, then the chemistry and the cooling. Then the same signals are sent again to the GPUs to perform the next time step until the simulation is completed. If required, the data are transfered back on the host to write the snapshot on the disk but such situation occurs only once in a while (typically once every 5000 time steps in our case). In such a procedure, only the host is aware of the fact that a time evolving calculation is performed but only the GPU does actual calculations. Furthermore, the global memory can be as big as 4GB on current devices and is used to store the data (like e.g. a large grid of values). This memory space is usually sufficient but slow in access. On-chip memory also exist, with fast access, but is usually small (of the order of 16 kB) and more importantly, still requires an access to the slow video memory to be filled. Therefore if possible, any memory access should lead to a significant `number crunching' in order to make these memory transfers worthwhile. {\tt ATON} fulfills this requirement quite easily since every time a cell is accessed (containing an energy density, flux, temperature, ionization fraction and baryon density), the cell is fully updated and requires a transport calculation and the resolution of the ionization and energy balance equations which are quite demanding in terms of arithmetic intensity.

Regarding the hierarchy of tasks, GPUs are efficient in performing tasks (or execution \emph{threads}) in parallels which are: 
\begin{itemize}
\item \emph{independent}. If a given thread has to wait for the completion of another one to perform its calculation (e.g. in a naturally sequential algorithm like a reduction \citet{Sengupta:2007:SPF}) or if two threads try to update simultaneously the same value (in e.g. histogramming calculations, \citet{1560857}), specific algorithmic techniques must be employed to keep the parallelism efficient. On the other hand, if calculations do not interfere with each other then porting these tasks on gpu architecture is usually quite easy. 
\item \emph{predictable}. Tasks can be unpredictable in their operations (if-else branches) or in their memory accesses. The former lead to divergence between threads where their execution tracks are executed sequentially by the GPUs thus reducing the efficiency of parallelism. If divergences are limited to exceptions (i.e. have a small chance to happen) and are hidden in intensive calculations their impact remain small. The latter lead to non coalescent and non aligned memory accesses which greatly impact the performances. 
\item \emph{compact}. A task is compact if it uploads data in a compact region in memory. Again, compact calculation leads to coalescent memory accesses which greatly improves the acceleration of the calculation. 
\end{itemize}
It turns out that {\tt ATON} possesses these three qualities. To demonstrate it, let us first recall that the radiation transport equations can be written in a generic manner as: 
\begin{equation}
\frac{du}{dt}+\frac{dF(u)}{dx}=S, 
\end{equation}
where $u$ is a set of conserved variables (energy density and flux in the case of radiative transfer), $F(u)$ the associated flux, and $S$ is a generic source term. We considered a 1D transport for simplicity. It translates into: 
\begin{equation}
\frac{\tilde u^{p+1}_i-u^p_i}{\Delta t}+\frac{ F^{p}_{i+1/2}(u)-F^{p}_{i-1/2}(u)}{\Delta x}=S_{i}^p, 
\end{equation}
when one considers an explicit finite difference (FD) scheme in order to update $u$ at position $i$ at time $p+1$. Usually the intercell flux can be exactly solved or approximated using the values in the neighboring cells through an operator $g$ and for instance: 
\begin{equation}
F_{i+1/2}=g(i,i+1/2). 
\end{equation}
Moreover the chemistry/temperature updates plus the effect of absorption can in our implementation be formally written as 
\begin{equation}
(u^{p+1}_i, x^{p+1}_i, T^{p+1}_i)=\Phi(\tilde u^{p+1}_i, x^{p}_i, T^{p}_i), 
\end{equation}
where $T$ and $x$ stand for the temperature and ionized fractions while $\tilde u^{p+1}_i$ stand for the conserved variables updated after the transport. As one deals with grid-based structures it is natural to assign a thread to the update of a specific cell. From there it can be seen that {\tt ATON} is well suited for GPU parallelism according to the three qualities listed before~: 
\begin{itemize}
\item \emph{independence~:} all these calculations are explicit~: the only intermediate results needed are the transport-updated $\tilde u^{p+1}_i$ and it is a local value. All the other inputs are initial state values which do not require communications during the calculation per se. As a consequence all the cell updates (and therefore the threads) are independent and the overall procedure is free of threads collisions or sequential calculations where one thread has to wait the completion of one or several other tasks. 
\item \emph{predictability~:} here the calculations are at least `memory-predictable'. Updating a given cell requires data in a region which is known by advance, i.e. the updated cell plus its 6 neighbors in 3D. Operation branching occurs in the cooling and chemistry calculations and has some impact on the performance (see the the subsequent analysis). 
\item \emph{compacity~:} again the calculation requires a 7 cells stencil for a single thread which is quite compact and allows to enforce the coalescent memory accesses, as shown hereafter. 
\end{itemize}

\begin{figure}
[htbp] \centering 
\includegraphics[width=0.7\columnwidth]{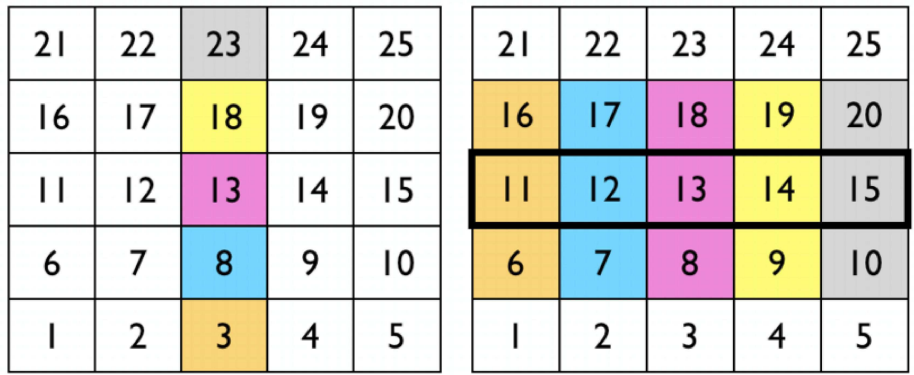} \caption{Comparison of two finite differences (FD) strategies on a 2D field in memory. The sequence indicates the actual organization of cells in memory, the coalescent direction. We consider the case where the FD should be performed along the non-coalescent direction. Each color represent the location to be computed by a thread. \emph{Left~:} `vertical strategy' where the threads are arranged along the FD direction. \emph{Right~:} `horizontal Strategy' where the threads are arranged perpendicular to the FD direction. The `horizontal strategy' maximizes the performance of the GPUs due to coalescent memory accesses. See main text for details.} \label{fig:plot_fd} 
\end{figure}

Finally, these devices can easily be used at full power if two properties are satisfied during the calculation: the data in global memory should be accessed in a \emph{coalescent} and \emph{aligned} fashion. Figure \ref{fig:plot_fd} allows us to explain the coalescence in details assuming a 2D calculation. The data is accessed in a coalescent fashion if a serie of threads reads data which are organized in a sequential manner in the memory. In Figure \ref{fig:plot_fd}, a 2D field is physically stored in memory as a 1D sequence listed by numbers 1 to 25. If a sequence of threads (shown colored) is set in such a way that threads access the data `vertically' (left scheme), they physically access data which are separated by jumps of 5 units~: such a strategy is non coalescent. Conversely if the sequence of threads is organized `horizontally' (bold border in the right scheme), they access to data which are physically next to each other, i.e. in a coalescent fashion. All the threads in {\tt ATON} are arranged following this strategy in order to enforce coalescence. For example the chemistry/cooling step is performed with threads along the physically coalescent direction. The radiation transport step is slightly more difficult to set up as it involves a finite difference along all the directions: 
\begin{equation}
\frac{\tilde u^{p+1}_{i,j}-u^p_{i,j}}{\Delta t}+\frac{ F^{p}_{i+1/2,j}(u)-F^{p}_{i-1/2,j}(u)}{\Delta x}+\frac{ F^{p}_{i,j+1/2}(u)-F^{p}_{i,j-1/2}(u)}{\Delta y}=S_{i,j}^p. 
\end{equation}
For the finite difference performed along the coalescent direction, the coalescence is naturally satisfied. In order to avoid multiple access to the same data by neighboring threads, the coalescent values are uploaded in shared (on chip) memory once and calculations are performed from this shared memory. For the finite difference performed along the non-coalesced direction (vertical in Figure \ref{fig:plot_fd}), a naive strategy would have been to upload the vertical values in shared memory (left column), i.e. along the direction of the finite difference. However this would imply non coalescent access. The correct way to deal with this finite difference is shown on the right panel of Figure \ref{fig:plot_fd}. First the threads should be organized along the coalescent direction (shown colored). Then all threads upload the data `above' the region to update (shown with a bold line) in shared memory along the coalescent direction. The same is done for the data `below' the region to update. Finally the finite difference can be performed. From our experience, switching from non coalescent to fully coalescent strategies can improve the performance of the GPU calculation by factor of 10 to 100. It should be said that such a `trick' is not specific to GPU-based calculation but given the high parallelization of the devices such an optimization has a more dramatic impact on their performances compared to usual scalar processors. 

\emph{Aligned access} is more specifically related to the hardware used. Typically, the data should be accessed in sets of 64, 96 or 128 words which is usually satisfied using thread configurations which rely on powers of two. A additional constraint is that the range of memories accessed by these sets should be aligned with `preferential' memory addresses, usually multiples of 16. When dealing with arrays with dimensions equals to powers of two, any access of sets of 64, 96 or 128 words will be automatically aligned. Non aligned access will result in multiple memory queries on aligned addresses in place of a single one. It turns out that such situation is quite common as boundary conditions usually add a layer of data around the actual computational volume making e.g. a $128\times 128 \times 128$ cube a $(128+2)\times (128+2)\times(128+2)$ cube, which breaks the alignment. We circumvent this by making the boundary layer larger than required by the code since we are not limited by memory (e.g. $128\times 128 \times 128$ cube becomes a $(128+32)\times (128+32)\times(128+32)$ cube). Typically an additional factor of 2-3 of acceleration can be achieved by enforcing alignment.

\begin{figure}
[htbp] \centering 
\includegraphics[width=.75\columnwidth]{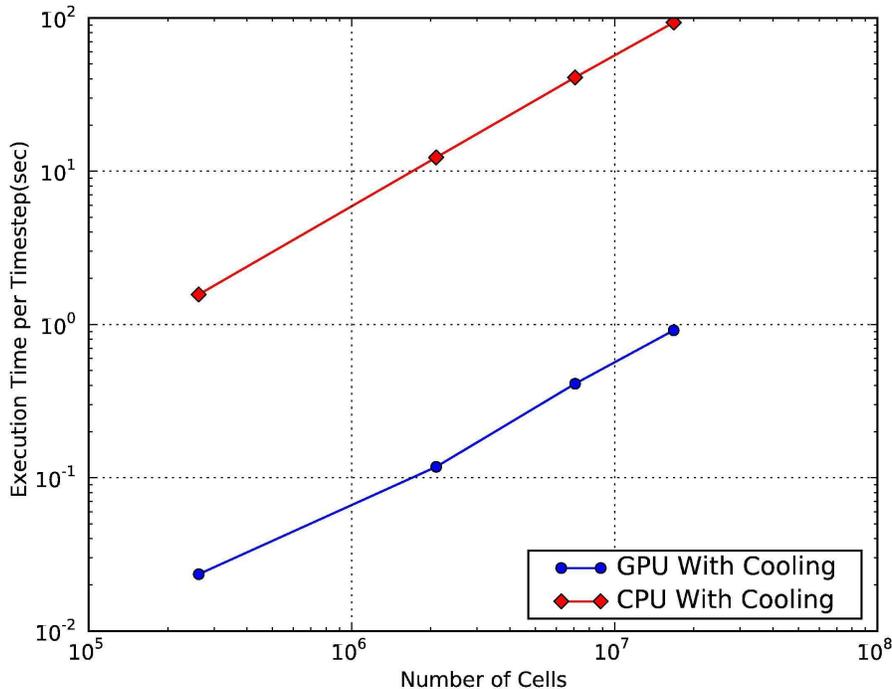} \caption{Average time steps duration for the cosmological test of the comparison project for CPU and GPU, at different resolutions. The GPU and the CPU are roughly equivalent in terms of generation.} \label{fig:plot_execcool} 
\end{figure}

As an illustration of the computing abilities of GPUs, we show in Fig. \ref{fig:plot_execcool} the average duration of a time step of a radiative transfer post-processing performed on the cosmological test of the comparison project. The same test has been performed at several resolution and executed on a Opteron 2.2 GHz and a GeForce 8800 GTX, which are comparable in terms of generation (2005-2006). A significant acceleration close to 80 was observed on GPU compared to a monocore run on CPU. Both calculations were performed using single float precision and no difference could be seen between the calculations at such precision. 

\begin{figure}
[htbp] \centering 
\includegraphics[height=3in]{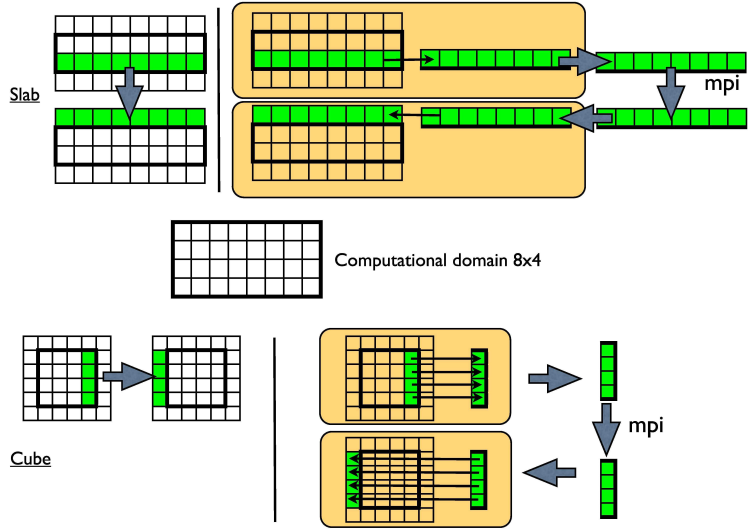} \caption{Two communication strategies for multi-gpu calculations. The coalescent direction is assumed to be the horizontal one. \emph{Top}~: slab-based communication. \emph{Bottom}~: cube-based communication. For the cube-based decomposition, some communications involve gathering and dispatching data in a non coalescent manner. The cube-based technique has nevertheless been chosen for {\tt ATON} to assess large problems, to reduce the shared memory usage and in the prospect of coupling {\tt ATON} to integrators with cube-based decomposition.} \label{fig:gpu} 
\end{figure}

\begin{figure}
[htbp] \centering 
\includegraphics[width=0.4\columnwidth]{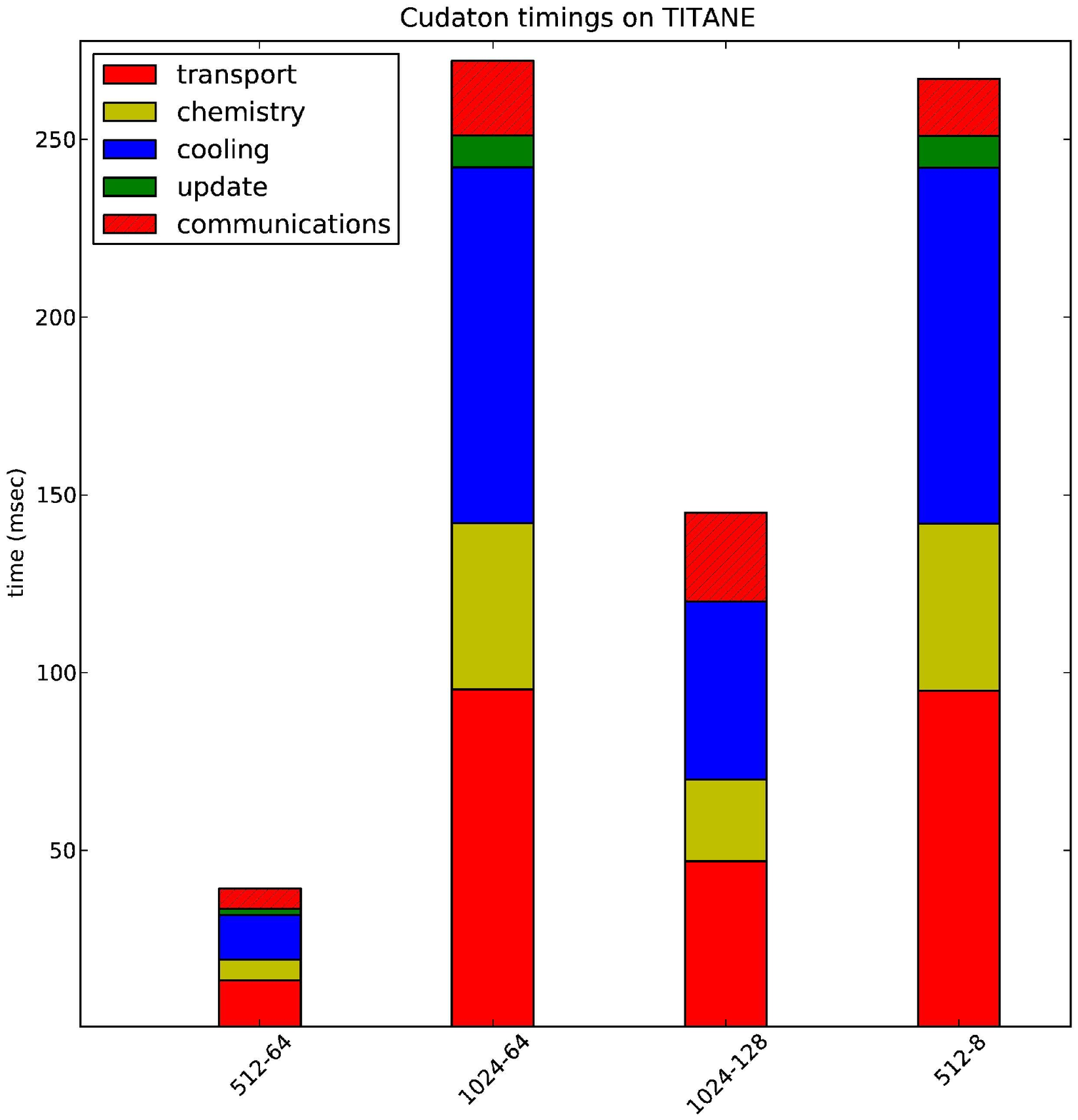} 
\includegraphics[width=0.4\columnwidth]{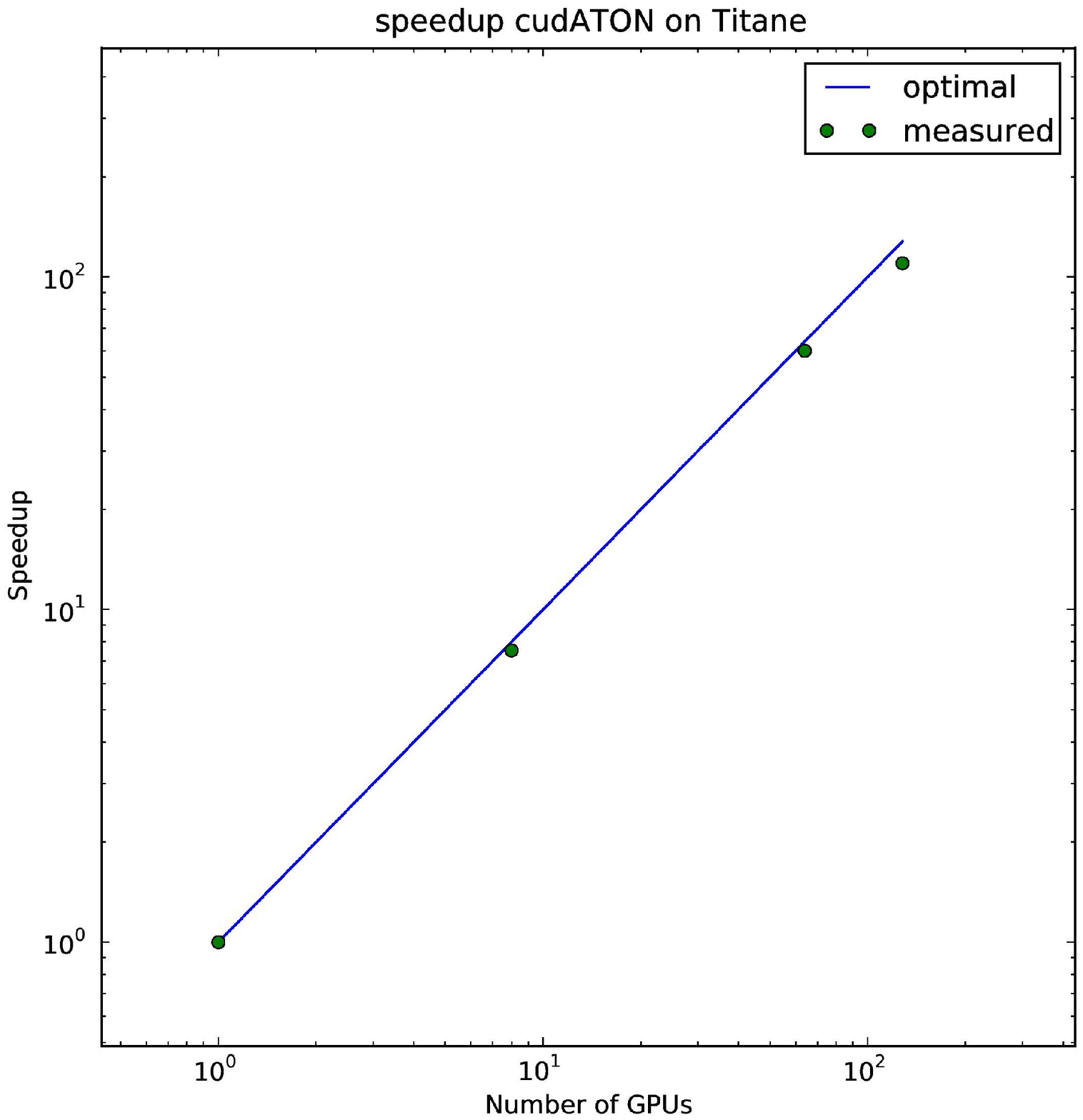} \caption{Timings
  of multi-GPU calculations. \emph{Left}: average duration of a time step for
  a typical cosmological field used in this work and for several parallel configurations. The first integer stand for the total number of cells along one direction and the second stands for the number of GPUs. For example 512-8 means a $512^3$ radiative transfer calculation distributed over 8 GPUs. \emph{Right}: Acceleration as a factor of the number of GPUs compared to a mono-GPU calculation, the dot stand for the actual measurement while the straight line stand for the perfect acceleration trend. Measurements were performed on the Titane supercomputer (CCRT-CEA) using Tesla C1060 GPUs.} \label{fig:plot_timing-titane} 
\end{figure}

{\tt ATON} is able to run on configurations with multiple graphical devices. It implies that GPUs should communicate in order to exchange boundary conditions. This is simply done by adding an MPI layer over the GPU inner parallelization. Once a GPU has updated its subgrid the following sequence happens at each time step: 
\begin{itemize}
\item A GPU-buffer is created on each GPU to gather the data to be passed at each time step (here namely the radiative energy and fluxes) 
\item this GPU-buffer is transmitted to the host into a CPU-buffer 
\item the CPU-buffers are exchanged using regular MPI-based instructions 
\item the updated CPU-buffers are transmitted to the GPUs into the GPU-buffer 
\item the data inside the buffer are distributed into the correct radiative variables. 
\end{itemize}
Considering the coalescence and alignment constraints depicted above, the natural parallelization for multi-gpu calculations is `slab-based' and for example a 512x512x512 calculation would be divided in 4 calculations 512x512x128 on 4 GPUs. The reason is that at each time step, the radiative energy and fluxes at the boundaries should be passed to the neighboring GPU by collecting them into a buffer and a slab-based configuration implies that the collection is performed in a coalescent manner (and the distribution as well). However we choose to stick to sub-cube based parallel configuration in the prospect of coupling with N-Body+hydro integrators (such as {\tt RAMSES}) which parallel configuration is closer to `sub-cube' segmentation than `slab-based' ones. Furthermore, the slab-based decomposition cannot be naively applied for large problems because of hardware limitation such as the amount of memory per kernel (16KB) or the number of threads per block (512). Conversely, it implies that non-coalescent access are performed during the gathering/distribution phases (see Fig. \ref{fig:gpu}). Finally it should be noted that such communications require systematic transfers between the hosts and the GPUs through the PCI bus, which act therefore as a bottleneck in the communications. In Fig. \ref{fig:plot_timing-titane} we present the average duration of the time steps for several multi-gpu configuration and problem sizes and the acceleration as a function of the number of GPUs. Even though the implemented parallelization is simple, the speedup trends is quite optimal and the amount of time spent into the communication remains reasonable at levels of 10-15$\%$. This number is quite important by standards of parallel high performance computing but compared to an initial acceleration of a few tens (compared to the CPU), this overhead remains small enough to assess large problems.

\end{document}